\documentclass{aa}

\usepackage[version=4]{mhchem}
\usepackage[varg]{txfonts}
\usepackage{graphicx}

\usepackage[colorlinks,bookmarks]{hyperref}
\usepackage{color}
\definecolor{linkblue}{rgb}{0,0,0.8}
\definecolor{linkgreen}{rgb}{0,0.5,0}
\hypersetup{linkcolor=linkblue,citecolor=linkblue, urlcolor=linkblue}

\newcommand{\be}{\begin{equation}}
 \newcommand{\ee}{\end{equation}}

\usepackage{booktabs}

\def\hzero    {{\rm{H^0}}}
\def\hezero    {{\rm{He^0}}}
\def\hi    {{\rm{H \, \textsc{i}}}}

\def\hisub {\rm{H \scriptscriptstyle I}}
\def\hiisub {\rm{H \scriptscriptstyle II}}

\def\nhi {n_{\hisub}}

\def\nhii {n_{\hiisub}}
\def\Siplus {\rm{Si}^+}
\def \ciiline {\rm{C \, \textsc{ii} \, \lambda 1334}}

\def \oilinesiii {\rm{O \, \textsc{i} \, \lambda 1302}}

\def \siii {\rm{Si \,  \textsc{ii}}}

\def \siiiline {\rm{Si \,  \textsc{ii} \, \lambda} 1260}
\def \siiilineb {\rm{Si \,  \textsc{ii} \, \lambda} 1526}
\def \siiidoublet {\rm{Si \,  \textsc{ii}} \, \lambda \lambda 1190, 1193}

\newcommand{\angstrom}{\mbox{\normalfont\AA}}
\def\Msun {\rm{M}_{\rm{\odot}}}
\def\sfr {\Msun \, \rm{yr}^{-1}}
\def\Zsun {\rm{Z}_{\rm{\odot}}}
\def\kms {\rm{km} \, \rm{s}^{-1}}                           % km/s
\def\fesc {f_{\rm{esc}}}
\def\fesclimit {f_{\rm{esc}}^{900}}
\def \ndust {n_{\mathrm{dust}}}
\def \sixteenth{16^{\rm{th}}}
\def \eightyfourth{84^{\rm{th}}}

\usepackage{natbib}
\bibpunct{(}{)}{;}{a}{}{,} % to follow the A&A style

\begin{document}

% \thanks{} \fnmsep 
\title{The {\sc sphinx} public data release}

\subtitle{II. Using low-ionisation absorption lines and dust attenuation to predict Lyman continuum escape} 

\author{Valentin Mauerhofer
  \inst{1}\fnmsep\thanks{\email{val.mauerhofer@gmail.com}}
  \and J\'er\'emy Blaizot \inst{2}
  \and Thibault Garel \inst{3}
  \and Anne Verhamme \inst{3}
  \and Simon Gazagnes
  \and Josephine Kerutt \inst{1}
  \and Leo Michel-Dansac \inst{4}
  \and Kaelee S. Parker \inst{5,6}
  \and Joakim Rosdahl \inst{2}
  \and Alberto Saldana-Lopez \inst{7}
  \and Maxime Trebitsch \inst{8}
  \and Taysun Kimm \inst{9}
  \and Pierre Ocvirk \inst{10}
  \and Romain Teyssier \inst{11}}

 \institute{Kapteyn Astronomical Institute, University of Groningen, PO Box 800, 9700 AV Groningen, The Netherlands
 \and Centre de Recherche Astrophysique de Lyon UMR5574, Univ Lyon1, ENS de Lyon, CNRS, F-69230 Saint-Genis-Laval, France
 \and Observatoire de Genève, Université de Genève, Chemin Pegasi 51, 1290 Versoix, Switzerland
 \and Aix Marseille Univ., CNRS, CNES, Laboratoire d’Astrophysique de Marseille, F-13388 Marseille, France
 \and Department of Astronomy, The University of Texas at Austin, 2515 Speedway, Stop C1400, Austin, TX 78712, USA
 \and Cosmic Frontier Center, The University of Texas at Austin, Austin, TX 78712, USA
 \and Department of Astronomy, Oskar Klein Centre, Stockholm University, 106 91 Stockholm, Sweden
 \and LUX, Observatoire de Paris, Université PSL, Sorbonne Université, CNRS, 75014 Paris, France
 \and Department of Astronomy, Yonsei University, 50 Yonsei-ro, Seodaemun-gu, Seoul 03722, Republic of Korea
 \and Observatoire Astronomique de Strasbourg, Université de Strasbourg, CNRS UMR 7550, 11 rue de l’Université,
67000 Strasbourg, France
 \and Department of Astrophysical Sciences, Princeton University, Peyton Hall, Princeton, NJ 08544, USA}

\date{}

\abstract
 {Low-ionisation state (LIS) metal absorption lines, such as $\siiilineb$, are widely used to trace the properties and dynamics of the interstellar medium (ISM) in galaxies. These lines provide crucial insights into galaxy evolution, including feedback mechanisms, metal enrichment, and the escape fraction of ionising photons ($\fesc$) during the epoch of reionisation.}
{We expand our understanding of LIS absorption lines as diagnostic tools for ISM properties and $\fesc$. Using the high-resolution {\sc sphinx}$^{20}$ cosmological radiation-hydrodynamics simulation, we generated a comprehensive synthetic dataset of LIS absorption lines and tested their predictive power for $\fesc$ in star-forming galaxies.}
 {Synthetic ISM absorption lines, focusing on $\siiiline$ and $\siiilineb$, were computed with the radiative transfer code {\sc rascas}, incorporating resonant scattering of photons, fluorescent emission, and interactions with dust grains. The simulated data enhance the public {\sc sphinx}$^{20}$ dataset with high-resolution LIS lines for the full 1380 galaxies and ten viewing angles per galaxy. We analysed correlations between line properties (width, depth, and Doppler shift), dust attenuation, and $\fesc$, extending previous single-galaxy studies to a statistically significant mock galaxy sample. We also tested our predictions on observed data using the LzLCS and CLASSY surveys.}
 {We found a strong correlation between the dust-corrected residual flux of $\siiilineb$, $\tilde{R} \equiv \rm{R_{flux}^{1526}} \cdot 10^{-0.4A_{1500}}$, and $\fesc$. More precisely, we found $\fesc \approx 1.041\tilde{R}^{1.887} - 0.002$, with an average absolute error of 0.02. When we applied observational conditions, the error increased, but the escape fraction was still well recovered. In particular, the measurement of residual fluxes required a very high spectral resolution, and the dust attenuation is not directly observable. We show by applying common tools for fitting the spectral energy distribution to our mock data that the inferred dust attenuation is often far from the correct value, with a tendency to underestimate the attenuation when the effect of dust is strongest.} 
 {Our results demonstrate that the residual flux of $\siiilineb$ is a powerful predictor of the escape fraction of ionising photons when it is corrected for dust. The spectra, line measurements, and escape fraction values used in this work are made publicly available.}

\keywords{radiative transfer -- line: formation -- dark ages, reionisation, first stars -- ultraviolet: galaxies -- galaxies}
\titlerunning{{\sc sphinx}$^{20}$ Absorption Lines Data Release}
\authorrunning{V. Mauerhofer et al.}
\maketitle 

%%%%%%%%%%%%%%%%%%%%%%%%%%%%%%%%%%%%%%%%%%%%%%%%%%%%%%%%%%%%%%%%%%%%%%%%%%%%%%%%%%%%
\section{Introduction}

The interstellar medium (ISM) in star-forming galaxies is a dynamic environment shaped by diverse processes, including star formation, stellar feedback, galactic winds, and gas accretion. Understanding the factors driving its evolution over time is critical for refining our models of galaxy formation and evolution. Observationally, this requires accurate methods for inferring the properties of the ISM. In this context, low-ionisation states (LIS) of metals, such as $\rm{C}^+$ or $\Siplus$, can serve as powerful tracers of neutral and low-ionisation ISM. When seen in absorption against the stellar continuum of galaxies (a so-called down-the-barrel observation), the widths, depth, asymmetry, shifts, and overall complexity of lines such as $\ciiline$ or $\siiiline$ encode geometric and dynamic information about the ISM \citep[e.g.][]{Steidel10}.

The LIS absorption lines are relevant for measuring outflow velocities and inferring mass outflow rates, as has been shown, for example, in \cite{Shapley03, Heckman15, Xu23_classy}. They were also used to study the metal enrichment of the ISM \citep[see, e.g.][]{James14, Chisholm18, James18}. Above all, LIS lines were used in the context of reionisation as a tracer of the escape fraction of ionising photons ($\fesc$). This crucial property of galaxies shows to which extent they contribute to ionising their surrounding intergalactic medium (IGM). 
Over the past two decades, several ground-breaking detections of Lyman-continuum (LyC) leakage have been achieved in galaxies at low and intermediate redshifts, providing key insights into how ionising photons escape from their ISM \citep[e.g.][]{Bergvall06,Vanzella10,Vanzella15,Vanzella18,Leitet13,Borthakur14,Izotov16a,Izotov16b,Izotov18b,Izotov21,Leitherer16,Puschnig17,Wang19,LzLCS_Flury_1,Marques-Chaves22,Rivera-Thorsen22,Saxena22}. To understand the reionisation era, it is essential to perform similar measurements at higher redshifts, but this is prevented by the increasing opacity of the IGM to ionising radiation \citep[e.g.][]{Fan06, 2018PASJ...70S..13O, Wise19}. Consequently, indirect tracers of LyC escape are required. Among these, LIS absorption lines have long appeared as promising diagnostics, since their strength and residual flux directly probe the distribution and covering fraction of neutral gas along the line of sight (LOS) \citep[e.g.][]{Heckman11, Erb15, Chisholm18, Gazagnes18, Steidel18}. Recently, the advent of large spectroscopic surveys such as the Low-z Lyman Continuum Survey \citep[LzLCS: ][]{LzLCS_Flury_1, LzLCS_Flury_2} has enabled statistical studies of these relations, revealing robust connections between $\fesc$ and the depth or residual flux of LIS lines \citep{Saldana-Lopez22}.

These empirical relations have motivated efforts to interpret the LIS line properties in terms of the physical conditions of the ISM and the mechanisms regulating LyC escape. Different modelling approaches have been explored to extract galaxy properties from observed LIS absorption lines. The picket-fence model \citep[e.g.][]{Reddy16, Chisholm18, Gazagnes18, Saldana-Lopez23} assumes that a fraction of stars is covered by optically thick gas, while the remaining stars face empty channels, or holes. In this model, the residual flux (i.e. the flux at the bottom of the line over the continuum flux) can be directly linked to $\fesc$. 
More sophisticated approaches have explored LIS line transfer in moving media and spherical anisotropic geometries using the Sobolev approximation or Monte Carlo simulations \citep[e.g.][]{Prochaska11,Scarlata15,Carr18,Garel24}.
While restricted to idealised geometries, these models are able to predict the effects of line infilling since they take multiple scatterings into account.
However, it is unclear whether the assumptions made by these models lead to an accurate representation of reality that can be used to infer physical properties of the ISM. In particular, they assume a unique central source, ignore absorption from the fine-structure level \citep{Mauerhofer21, Gazagnes23}, do not account for the Doppler effects emerging from the complex dynamics of gas, ignore the effect of satellite galaxies within the LOS, and oversimplify the complex star-gas-dust geometry (including the precise ionisation state of the gas and over-abundance of dust in cold dense regions).

This gap hampers our ability to interpret the LIS lines quantitatively and to use them as robust diagnostics of the escape of ionising radiation. A self-consistent dataset of simulated absorption spectra, tied to well-defined galaxy properties, is therefore crucial.
To address this issue, we took advantage of the rich ensemble of high-redshift star-forming galaxies present in the {\sc sphinx}$^{20}$ simulation \citep{Sphinx20}, which is a high-resolution cosmological radiation-hydrodynamics simulation that follows the evolution of tens of thousands of galaxies down to $z=4.64$. 
We enhance the public dataset of the {\sc sphinx}$^{20}$ simulation \citep{Sphinx20_release} by including high-resolution synthetic ISM absorption lines, focusing on $\siiiline$ and $\siiilineb$. We chose these lines because they are among the strongest low-ionisation transitions and remain observable at high redshift, unlike lines shortward of $\rm{Lyman}$-$\alpha$ such as $\rm{O \, \textsc{i} \, \lambda 1039}$ or $\siiidoublet$. While other redder lines such as $\oilinesiii$ or $\ciiline$ are also strong, the presence of nearby fluorescent emission or of other absorption lines complicates the interpretation. With the chosen lines, we built a catalogue of realistic LIS absorption lines, which can then be associated with non-observable galaxy properties based on the plethora of data already presented in the data release \citep{Sphinx20_release}. This can be used as tests for current and future models of $\siii$ absorption (and fluorescent) lines, and also for a direct comparison to observations of star-forming galaxies (as in e.g. \cite{Gazagnes23, Gazagnes24}).

\defcitealias{Mauerhofer21}{M21}

As an example application of our catalogue of lines, we extend the work of \citet[][hereafter, M21]{Mauerhofer21}, where we found a correlation between the residual flux of LIS lines, the dust attenuation, and $\fesc$ in a single galaxy taken from a \textsc{sphinx}-like cosmological zoom simulation at different times (from $z=3.2$ to $z=3$) and viewing angles. 
Given the importance of identifying reliable indirect tracers of LyC escape, we test here whether these relations persist across a broader population of simulated galaxies and if they might thus be applied to high-redshift observations.
We also discuss the obstacles when our predictions are applied on real data, and we test it with observed low-redshift star-forming galaxies using LzLCS and the COS Legacy Archive Spectroscopy Survey \citep[CLASSY: ][]{Berg22, James22}.

The paper is structured as follows. In Section \ref{sec:sim} we detail the simulation and our modelling techniques. In Section \ref{sec:fesc_siii1526} we show how well we can predict $\fesc$ using the line properties of $\siiilineb$ and the values of the UV dust attenuation. In Section \ref{sec:discussion} we discuss the limitations of our escape fraction predictions, in particular, we test how accurately the dust attenuation can be estimated via Spectral Energy Distribution (SED) fitting. We also compare our results with observations. Finally, we summarise our findings in Section \ref{sec:summary}.

%-----------------------------------------------------------------------
\section{Simulation and methods} \label{sec:sim}

In this section, we describe the new data, which we are making public. We then explain our procedure to generate mock absorption lines.

\subsection{Set of simulated galaxies}
We analysed virtual galaxies from the radiation-hydrodynamics simulation {\sc sphinx}$^{20}$, presented in \cite{Sphinx20}. {\sc sphinx}$^{20}$ is a cubic volume of (20 cMpc)$^3$ simulated using the \textsc{Ramses-RT} adaptive mesh refinement code \citep{Ramses, Joki2013}. It resolves haloes down to the atomic cooling mass of $3 \times 10^7\Msun$, as a result of its dark matter particle mass of $2.5 \times 10^5 \Msun$. The code includes on-the-fly radiative transfer (RT) in two bins of hydrogen-ionising photons using the M1 method \citep{Joki2013}.
This volume was selected as the most representative of 60 cosmological initial conditions in order to obtain an average reionisation history. The gas resolution reaches 10 pc in the densest regions of the ISM, which is crucial to resolve the escape of ionising radiation and the photochemistry of interstellar gas. In star formation, gas is converted into stars in cells with a gas density higher than 10 $\rm cm^{-3}$, a locally turbulent Jeans length smaller than the cell width, and gas that is locally convergent \citep{Sphinx}. When a cell satisfies these conditions, its star formation efficiency per free-fall time is computed as a function of the local virial parameter and turbulence. The feedback from type II supernova (SNII) explosions is implemented as in \citet{Kimm15} as mechanical feedback. The SNII rate is artificially boosted by a factor of roughly four compared to a Kroupa initial mass function, which is necessary to suppress star formation and produce realistic high-redshift luminosity functions \citep{Sphinx20}.

We used the same subset of galaxies as in \cite{Sphinx20_release}, where all galaxies were selected with a star formation rate averaged over the previous $10 \, \rm{Myr}$ (SFR$_{10}$) higher than $0.3 \sfr$ in seven simulation snapshots at redshifts 10, 9, 8, 7, 6, 5, and 4.64. This resulted in a collection of 1380 galaxies, for which mock observations were produced along ten different directions of observation, isotropically distributed on the unit sphere. These galaxies span a range of stellar mass of $10^{6.5} \, \Msun$--$10^{10.5} \, \Msun$, an SFR of$_{10}$ $0.3 \, \sfr$--$80 \, \sfr$, and a gas-phase metallicity (mass weighted $12+\log(\rm{O/H})$) of $6.1$--$8.4$. Further data on the selected galaxy populations are presented in details in Figures 3-12 of \cite{Sphinx20_release}. We now supplement the {\sc sphinx}$^{20}$ data release with absorption lines of $\siiiline$ and $\siiilineb$, which were simulated as follows.

%-------------------------------------------------------------------------------
\subsection{Modelling absorption lines} \label{sec:abs_lines}

To model $\siiiline$ and $\siiilineb$, we followed the method described in \citetalias{Mauerhofer21} and \cite{Gazagnes23}. More precisely, we used {\sc rascas} \citep{Rascas} to propagate the stellar continuum from the stellar particles into the ISM and CGM of the simulated galaxies, until photons were either absorbed by dust or escaped from the virial radius. The continuum photons were sampled from the {\sc bpassv2.2.1} \citep{BPASS1,BPASS2} library (the same as in {\sc sphinx}$^{20}$), and distributed among the stellar particles, in proportion to their luminosities as photon packets. We used at least $10^6$ photon packets per galaxy, with a linear increase (proportional to stellar mass) up to a maximum of $10^7$ photon packets for the most massive galaxies. We confirmed that this was sufficient to obtain mock spectra with minimal numerical noise.

To set up the gas and dust medium into which the radiative transfer occurs, the density of silicon atoms in each cell was computed based on the hydrogen density and metallicity. We assumed solar abundance ratios scaled to the local metallicity $\rm Z_{cell}$: $\rm n_{Si} = n_{H} A_{Si} Z_{cell}/\Zsun$, where $\rm A_{Si}$ is the solar ratio of silicon atoms over hydrogen atoms ($3.24 \times 10^{-4}$), and $\Zsun$ is the solar metallicity (0.0139), both taken from \cite{Asplund21}. As in \cite{Gazagnes23}, we included a model of depletion onto dust grains, since a non-negligible fraction of silicon is not in the gas phase, available for absorbing stellar continuum photons, but is locked in dust grains. To do this, we followed the results of \cite{DeCia16} and \cite{Konstantopoulou22}, who measured the dust depletion of diverse galaxies and reported trends between galaxy metallicity and depletion factors. We applied these trends on a cell-by-cell basis, using the cell metallicity to derive the dust depletion factor of silicon. More quantitatively, we removed a fraction 
$$\delta_{\rm{Si}} \equiv 1-10^{-0.04-0.72\left(1.09+0.6 \log_{10}\left[\rm{Z_{cell}}/\Zsun\right] \right)}$$
of silicon from each cell. This naturally led to a slight decrease in the equivalent width of absorption and fluorescent emission.

We then computed the ionisation fraction of silicon to derive the density of $\Siplus$. To do this, we used \textsc{Krome}\footnote{\url{http://kromepackage.org}} \citep{2014MNRAS.439.2386G}, with a chemical network consisting of collisional rates from \cite{Voronov97}, recombination rates from \cite{BadnellRR}, and photoionisation rates from \cite{Verner96}. We fixed the ionisation fraction of hydrogen and helium to their simulation values, and we let the ionisation fraction of silicon evolve to equilibrium. For the photoionisation, we used the radiation field from the simulation for energies above 13.6 eV. However, neutral silicon atoms can also be ionised by photons at energies between 8.15eV and 13.6eV, which are not included in {\sc sphinx}$^{20}$. In \citetalias{Mauerhofer21}, we used a UV background taken from \cite{HM12} in each cell with $n_{\hi}<10^2 \, \rm{cm}^{-3}$. Instead, we now used the results of \cite{Katz22}, who restarted each simulation snapshot we use in this paper, freezing everything but the radiation, to propagate new radiation bins with energies below 13.6 eV. This provided an accurate sub-ionising radiation field in every cell of the simulation. We note that the LIS spectra are practically unchanged by this. 

\begin{figure}
  \resizebox{\hsize}{!}{\includegraphics{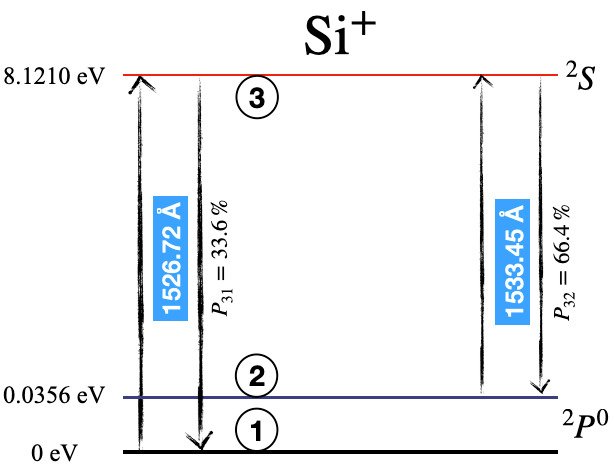}}
  \caption{Energy levels of the $\Siplus$ ion and transitions of $\siiilineb$. $P_{31}$ and $P_{32}$ are the probabilities that a $\Siplus$ ion in level 3 radiatively de-excites to level 1 or 2. 
  }
  \label{fig:SiII_structure}
\end{figure}

In our mocks, we also accounted for the fluorescence associated with $\Siplus$ transitions. Fluorescent emission occurs when an ion de-excites to the fine-structure level instead of the true ground state. We illustrate this in Fig.\ref{fig:SiII_structure}, which shows the different levels involved in the absorption and fluorescence of $\siiilineb$. The shape of the fluorescence is sensitive to transitions that start from the fine-structure level (level 2 in Fig.\ref{fig:SiII_structure}), which occurs because a fraction of $\Siplus$ atoms in the ISM is excited by collisions with electrons. We modelled this as in \citetalias{Mauerhofer21}, by computing this fraction of ions populating the fine-structure level using PyNeb\footnote{\url{https://pypi.org/project/PyNeb/}} \citep{Pyneb}, based on the electron density and temperature in each cell. This allowed us to take all the transitions shown in Fig.\ref{fig:SiII_structure} into account during the {\sc rascas} runs. For comparison, we also show the diagram for $\siiiline$ in Appendix \ref{sec:1260_plots}. It is slightly more complex, with an additional fine-structure of the upper level.

The dust opacity in every gas cell was computed assuming the implementation of \cite{Laursen09b} using the SMC law, as in \cite{Mauerhofer21, Sphinx20_release}. In this model, the optical depth of dust is the product of a pseudo-density and a cross section. The pseudo-density is
\be \label{eq:ndust}
\ndust = \frac{\rm{Z_{cell}}}{0.005} \times (\nhi + 0.01 \nhii).
\ee
For the cross-section, we used the results of \cite{Gnedin08} in the SMC case. This accounts for absorption and scattering events. For the albedo of dust grains, we used the values of \cite{Li01}, which are 0.338 for $\siiiline$ and 0.431 for $\siiilineb$. 
The direction in which a photon goes after scattering on dust was modelled with the Henyey-Greenstein function \citep{Henyey41}. This contains an asymmetry parameter $g$, also taken from \cite{Li01}. It is 0.591 for $\siiiline$ and 0.575 for $\siiilineb$.

After this setup, we computed the optical depth in each cell, and we took $\siiilineb$ as an example:
\be \label{eq:tau_cell}
\tau_{\rm{cell}} = \tau_{\siii \, \lambda 1526.7} + \tau_{\siii^{\star} \, \lambda 1533.4} + \tau_{\rm{dust}}.
\ee

The optical depth of a given line in a gas cell is computed as $\tau_{\rm{line}}=\sigma_{\rm{line}} \rm{N_{\Siplus}^{(*)}}$, where $\rm{N_{\Siplus}^{(*)}}$ is the column density of ionised silicon in the ground state (fine-structure level) along the path in the cell, and the cross section is 
\be
\sigma_{\rm{line}}(x,a) = \frac{\sqrt{\pi} e^2 f_{\rm{line}}}{m_e c} \frac{\lambda_{\rm{line}}}{b} \rm{Voigt}(x,a),
\ee
where $f_{\rm{line}}$ is the oscillator strength of the line, $\lambda_{\rm{line}}$ is the central wavelength, and $b$ is the Doppler parameter, which we explain below.
Additionally, the variable $a$ is defined by $a = A_{\rm{line}} \lambda_{\rm{line}} / (4 \pi b),$ where $A_{\rm{line}}$ is the Einstein coefficient of the line\footnote{All the line parameters are taken from the NIST atomic database \url{https://www.nist.gov/pml/atomic-spectra-database}}. The variable $x$ is a normalised wavelength shift from the line centre and is defined by $x = (c/b) (\lambda_{\rm{line}} - \lambda_{\rm{cell}})/\lambda_{\rm{cell}},$ where $\lambda_{\rm{cell}}$ is the wavelength of the photon packet in the frame of reference of the cell. We computed the Voigt function with the approximation of \cite{Smith_Colt}. 

When an interaction occured, one of the three channels of Equation \ref{eq:tau_cell} was randomly chosen with a probability $\tau_{\rm{channel}}/\tau_{\rm{cell}}$. When the photon packet is absorbed by either the ground state or fine-structure level of $\rm{Si}^+$, it is re-emitted in a direction randomly drawn from an isotropic distribution. When the absorption channel is $1526.7 \, \angstrom$, the photon packet is re-emitted either via the same channel, as resonant scattering, or via the $1533.4 \, \angstrom$ channel, which is fluorescent. The probabilities of the two channels are determined by the ratio of their Einstein coefficients, and they are shown in $\rm{Fig.} \, \ref{fig:SiII_structure}$. Similarly, when the photon packet is absorbed by channel $1533.4 \, \angstrom$, it can be re-emitted at $1526.7 \, \angstrom$ or at $1533.4 \, \angstrom$.

\begin{figure*}
  \resizebox{\hsize}{!}{\includegraphics{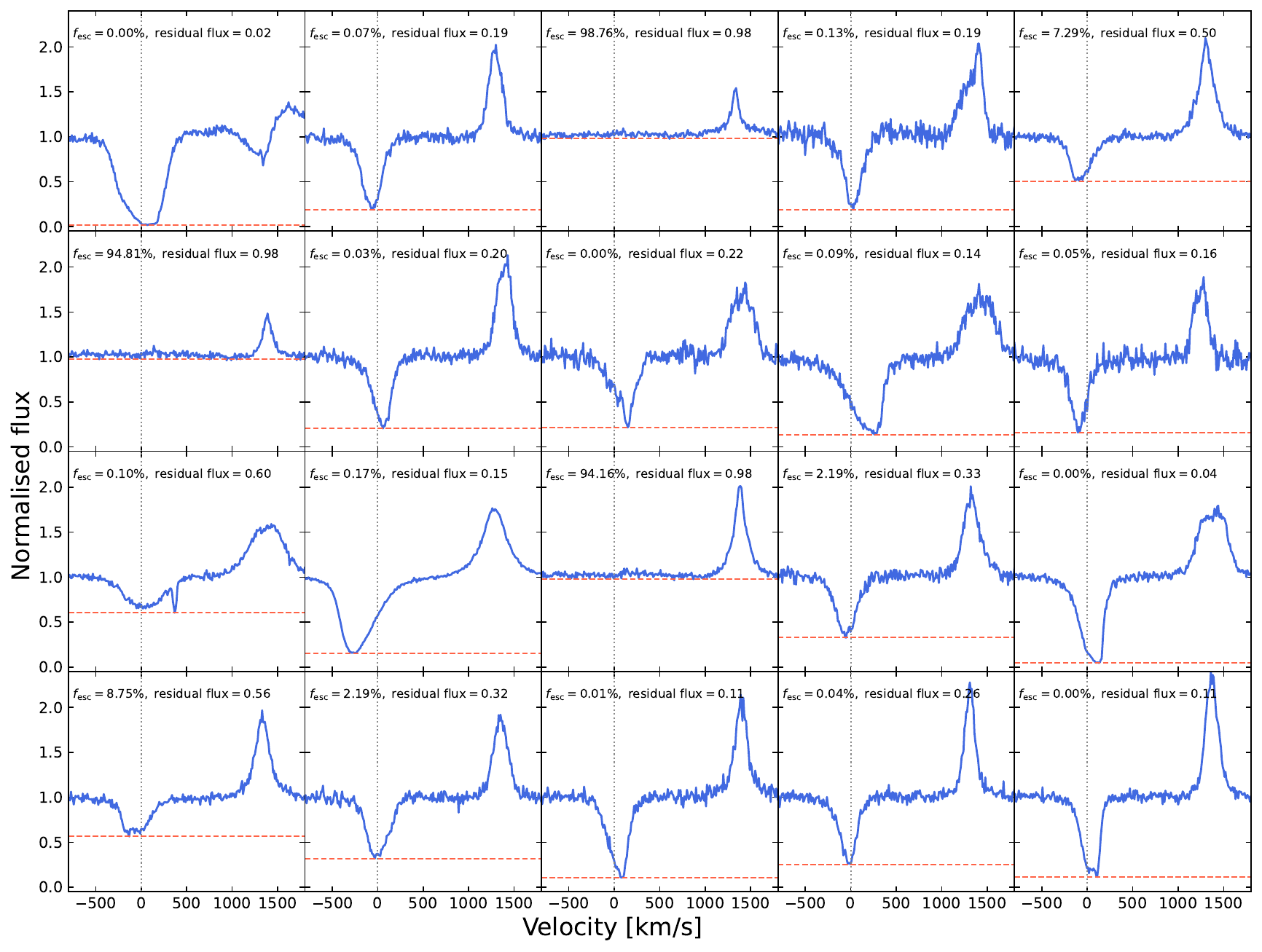}}
  \caption{Examples of $\siiilineb$ absorption and fluorescent emission profiles drawn from different haloes and LOS among our seven simulation snapshots. The spectra were selected using KMeans clustering to span a diverse range of spectral properties, including the equivalent width of absorption, residual flux, velocity centroid, and fluorescence strength. Each panel shows the normalised spectrum (blue) along with a dashed horizontal line at the minimum flux level, and a vertical dotted grey line at line centre. The red line highlights the value of the residual flux, which is indicated at the top of each panel, along with the corresponding $\fesc$.}
  \label{fig:mosaic}
\end{figure*}

We explain the Doppler parameter $b$, which was updated compared to \citetalias{Mauerhofer21}. It is a combination of thermal velocity and turbulent velocity $v_{\rm{turb}}$,
\be \label{eq:vth}
b = \sqrt{\frac{2 k_b T}{m_{\rm{ion}}} + v^2_{\rm{turb}}},
\ee
where $k_b$ is the Boltzmann constant, $T$ is the temperature, and $m_{\rm{ion}}$ is the mass of the interacting ion. In \citetalias{Mauerhofer21}, we used a global constant $v_{\rm{turb}}$ of 20 $\kms$. 
To avoid using this free parameter, we now computed a turbulent velocity in each cell based on the nine density-weighted velocity gradients of the cell (three axis directions for $v_x$, $v_y$, and $v_z$), which were computed from the properties of its direct neighbours. 
The resulting distribution of turbulent velocities has a volume-averaged value of $\sim 100 \,\kms$ and a density-averaged value of $\sim 25 \,\kms$, calculated within the virial radius of our simulated galaxies. Overall, this model of turbulence provides significantly broader absorption lines and fluorescent emission than the fiducial model of \citetalias{Mauerhofer21}.

%......................................................................................
\subsection{Computing line properties from spectra} \label{sec:line_properties}

The output absorption line spectra have a resolution of 10 $\kms$ between $-1500 \kms$ and $v_{\rm{fluo}} + 1500 \kms$, where $v_{\rm{fluo}}$ is the velocity of the fluorescent emission for each line. We used an aperture diameter of 2$\arcsec$, which means that the physical size of the aperture depended on redshift ($\sim 8.5 \text{--} 13 \, \rm{kpc}$), but was always large enough to encompass the whole galaxy. To normalise the spectra, we divide them by the value of the continuum luminosity at their (flat) edges.
We then computed the equivalent width (EW$_{\rm{abs}}$) of the absorption lines by integrating the spectra where the normalised luminosity is below 1 in a region from $-1000 \kms$ to $+600 \kms$ around the wavelength of the absorption.  
We defined EW$_{\rm{abs}}$ to be positive. 
Then, the residual flux $R_{\rm{flux}}^{\rm{line}}$ of the lines was computed by taking the minimum flux of the normalised spectrum.
Finally, the centroid velocity $v_{\rm{cen}}^{\rm{line}}$ was computed by determining the velocity at which the absorption lines had half their EW. These measurements were made with the same method as in \cite{Gazagnes23}. 

To illustrate the range of spectral properties in our sample, we selected representative spectra from each snapshot using KMeans clustering on key line properties, including the absorption equivalent width, residual flux, velocity centroid, and fluorescence strength. This procedure ensured that the selected spectra captured the diversity present in the simulations. These examples are shown in Fig.~\ref{fig:mosaic}. The resulting spectra exhibit a variety of features: some lines are nearly symmetric (top left panel), while others are skewed (second row in the fourth column, or third row in the second column); the panel in the third row and first column shows an example of a profile with a secondary absorption feature; some LOS have no absorption; and the level of noise varies due to differing dust attenuation along the LOS.

\subsection{Computing the escape fraction of ionising photons}

We are interested in predicting the LyC escape fraction using LIS absorption lines. While the data release of \cite{Sphinx20_release} does contain values for $\fesc$ in ten viewing angles, it was computed only at a wavelength of $900 \AA$. 
We computed the escape fractions as in \citetalias{Mauerhofer21}. To summarise, we sampled the intrinsic continuum of our galaxies from $200\AA$ up to the Lyman limit and propagated it until the virial radius using {\sc rascas}. Photons interact with $\hzero$, $\hezero$, $\rm{He^+}$, and dust. The cross section of the interaction between ionising photons and $\hzero$ and $\rm{He^+}$ is computed analytically \citep{Osterbrock},
\be \label{eq:sigma_ionizing}
\sigma_{\hzero,\rm{He^+}}(\nu) = \frac{6.3\times10^{-18}\,\rm{cm^2}}{\rm{Z}^2} \left( \frac{\nu_0}{\nu} \right)^4 \frac{\exp \left(4-4\frac{\arctan(\epsilon)}{\epsilon} \right)}{1 - \exp\left( \frac{-2 \pi}{\epsilon}\right)},
\ee
where $\nu$ is the frequency of the photon in Hz, assumed to be high enough to be ionising, Z is the nuclear charge (1 for $\hzero$ and 2 for $\rm{He^+}$), $\nu_0$ is the ionisation threshold frequency ($13.6 \, \rm{eV}$ or $54.4 \, \rm{eV}$ over the Planck constant, for $\hzero$ and $\rm{He^+}$, respectively), and $\epsilon=\sqrt{\nu/\nu_0-1}$. For the cross section of the interaction with $\hezero$, we used the approximation of \cite{Verner96}. For the dust cross section, we used the same implementation as for the non-ionising part, which included an extrapolation for ionising wavelengths \citep{Gnedin08, Laursen09b}.

\begin{figure}
  \resizebox{\hsize}{!}{\includegraphics{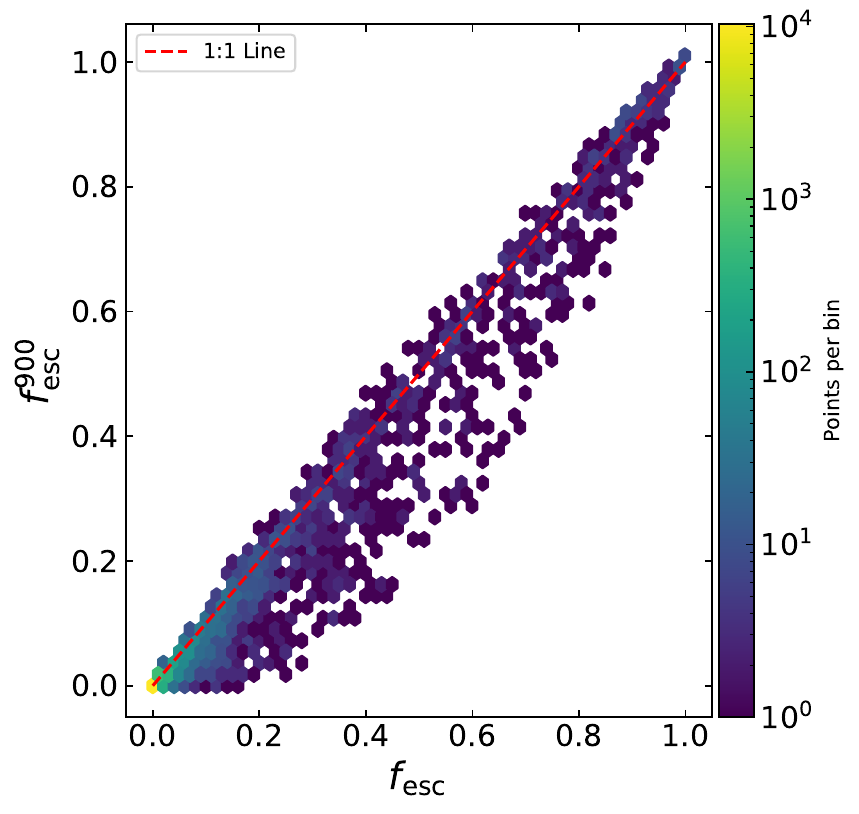}}
  \caption{Comparison of the escape fraction integrated over all ionising wavelengths on the x-axis and the escape fraction between $890 \angstrom$ and $910 \angstrom$ on the y-axis for all our simulated galaxies in the seven snapshots and in ten viewing angles.
  }
  \label{fig:fesc_fesc_900}
\end{figure}

Just as for the {\sc rascas} runs for the absorption lines, we used the peeling-off algorithm to measure the ionising continuum spectra in ten observation directions. The escape fraction is then simply the ratio of the integral of the escaping ionising continuum over the integral of the intrinsic ionising continuum. This provided us with the escape fraction of all ionising photons ($\fesc$) and with the escape fraction of photons around $900 \angstrom$, which we computed from $890 \angstrom$ to $910 \angstrom$ ($\fesclimit$). The former is important when considering the contribution of galaxies to reionisation, while the latter is useful for a comparison with observations, since most of the time, we only have access to the ionising spectrum near $900 \angstrom$. The two escape fractions are included in the new data release. We show in Fig.\ref{fig:fesc_fesc_900} that both escape fractions have relatively similar values, with differences up to about 0.25 at maximum. In most cases, $\fesc \ge \fesclimit$, since $\hi$ absorbs more photons at $900\angstrom$ than at lower wavelength (see Equation \ref{eq:sigma_ionizing}). The rarer cases when $\fesc < \fesclimit$ can be explained by LOS where hydrogen is ionised, letting $900\angstrom$ photons pass, but the presence of helium absorbs a fraction of lower-wavelength photons.

\medskip
Finally, we summarise our model in Table \ref{tab:comp} and compare it to the older model by \citetalias{Mauerhofer21}. We highlight all the novelties of the present work. All the data files are described in Appendix \ref{sec:release_content}, and they are publicly available (see the section with the data availability).

\begin{table*}[ht] \label{tab:comp}
\centering
\caption{Summary of novelties between \citetalias{Mauerhofer21} and this work.}
\begin{tabular}{p{4cm} | p{6cm} p{6cm}}
\hline\hline
\rule{0pt}{2.6ex} % adds space above the first row
 & M21 & This Work \\
\hline
Simulation used & Single zoom-in galaxy, three snapshots at $z\sim 3$, 1728 sightlines each & Full \textsc{sphinx20} volume  in seven snapshots at $4.6 \leq z \leq 10$ (1380 galaxies total, 10 sightlines each) \\[2pt]
Absorption lines & $\ciiline$ and $\siiiline$ & $\siiiline$ and $\siiilineb$ \\[2pt]
Turbulence & Fixed $v_{\mathrm{turb}} = 20~\mathrm{km\,s^{-1}}$ (tested $0$--$50~\mathrm{km\,s^{-1}}$) & Cell-by-cell turbulence from local velocity gradients (typical $25$--$100~\mathrm{km\,s^{-1}}$) \\[2pt]
Dust depletion & Neglected & Si depletion onto grains included, following \cite{DeCia16} and \cite{Konstantopoulou22} \\[2pt]
Sub-ionising radiation field & UV background from \cite{HM12} between 6 eV and 13.6 eV & Accurate inhomogeneous radiation bins from \cite{Katz22} restarts \\[2pt]
Data output & Not published & Public dataset of LIS spectra, $f_{\mathrm{esc}}$, and dust attenuation for all galaxies (see Data availability section) \\
\hline
\end{tabular}
\end{table*}

%%%%%%%%%%%%%%%%%%%%%%%%%%%%%%%%%%%%%%%%%%%%%%%%%%%%%%%%%%%%%%%%%%%%%%%%%%%%%%%%%
\section{Inferring $\fesc$ from $\siii$ absorption lines} \label{sec:fesc_siii1526}

To show one possible use of this new data release, we explore how well we can infer $\fesc$ using the properties of our LIS absorption lines,  such as the equivalent width of absorption, the residual flux, or the centroid velocity. 
Then, we explore whether using the dust attenuation factor as in \citetalias{Mauerhofer21} provides an accurate estimate of $\fesc$ when using a large number of galaxies with diverse properties rather than a single zoomed-in simulation.

%......................................................................................
\subsection{Correlations with escape fractions} \label{sec:correlations}

\begin{figure*}
  \resizebox{\hsize}{!}{\includegraphics{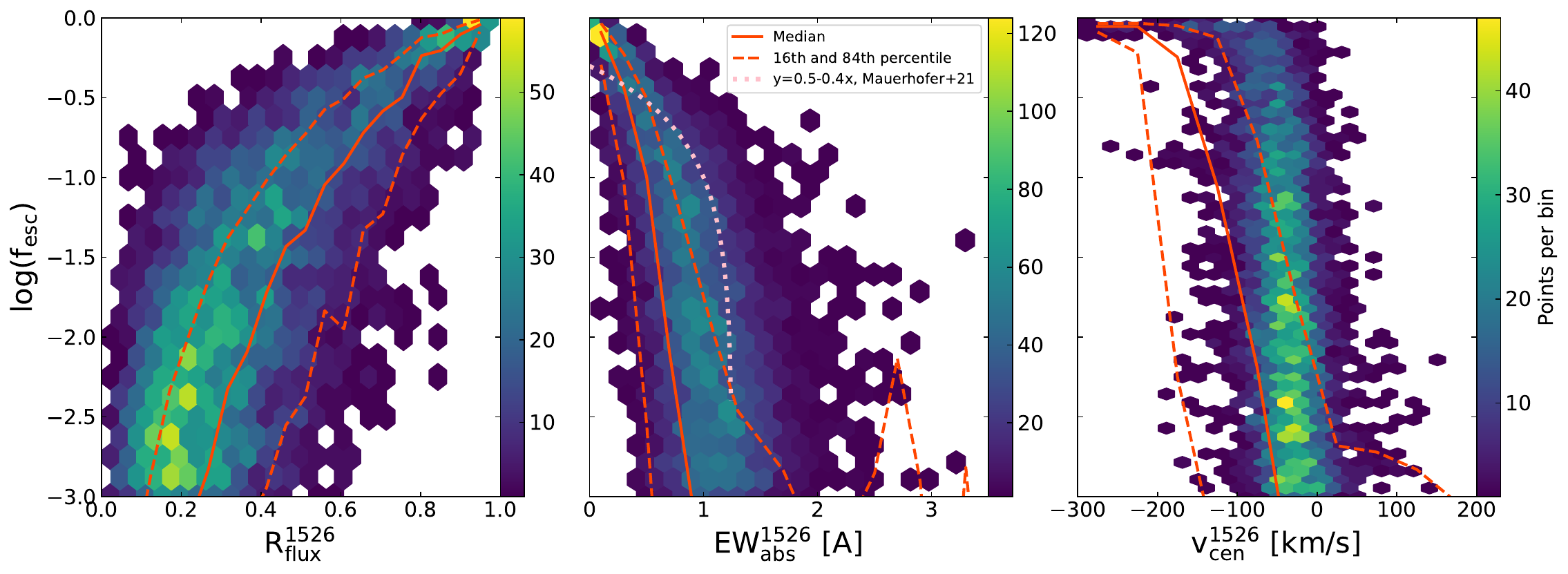}}
  \caption{Relations between the escape fraction of ionising photons and three $\siiilineb$ properties. The solid lines show the running median, and the dashed lines show the $\sixteenth$ and $\eightyfourth$ percentiles. These lines are affected by spectra with $\log(\fesc)<-3$, which are not displayed here. The left and middle panels show the residual flux and equivalent width of the absorption line, respectively. The dotted pink line shows the upper limit found in \citetalias{Mauerhofer21}. The right panel shows the centroid velocity of $\siiilineb$. }
  \label{fig:SiII_prop_fesc}
\end{figure*}

We now analyse whether some line properties correlate with the escape fraction of ionising photons. In Fig.\ref{fig:SiII_prop_fesc} we show the relations between the escape fraction and three $\siiilineb$ properties, that is, the residual flux, the equivalent width of the absorption, and the centroid velocity. In the following, we focus on this specific LIS line because we found that it yields better predictions than $\siiiline$. Some results using the latter line are shown in Appendix \ref{sec:1260_plots}, however.

There is a weak correlation between $\fesc$ and the residual flux of $\siiilineb$, as shown in the left panel of Fig.\ref{fig:SiII_prop_fesc}. Saturated lines, meaning lines with a residual flux below $\sim0.1$, almost all have an escape fraction below 10\%. Most of the spectra with a residual flux above $\sim0.75$, that is, with weak absorption features, display a large escape fraction, $\fesc \gtrsim 5\%$. Except for these edge cases, $\rm{R_{flux}^{1526}}$ does not predict $\fesc$. 

The middle panel displays a poor correlation between the escape fraction and the EW of absorption. The only constraint we can derive is that spectra with a large EW rarely have an escape fraction above a few percent (e.g. only $0.9\%$ of the spectra with EW$>1.5 \, \angstrom$ have $\fesc>5\%$). We also plot the upper limit that was found in \citetalias{Mauerhofer21} as a pink dotted line, and find that the {\sc sphinx}$^{20}$ galaxies do not follow this limit. This is mainly because in our larger sample, some galaxies have escape fractions above the maximum of the isolated galaxy studied in that paper, and some galaxies have larger equivalent widths of the absorption. 

Finally, the right panel shows no correlation between the escape fraction and the centroid velocity, which agrees with the results from \cite{Chisholm17}. The values below $\sim -150 \kms$ are uncertain, since they are associated with spectra having little absorption, for which it is hard to measure the velocity of the line due to noise.

%......................................................................................
\subsection{Dust correction to improve $\fesc$ predictions} \label{res:dust}

\begin{figure}
  \resizebox{\hsize}{!}{\includegraphics{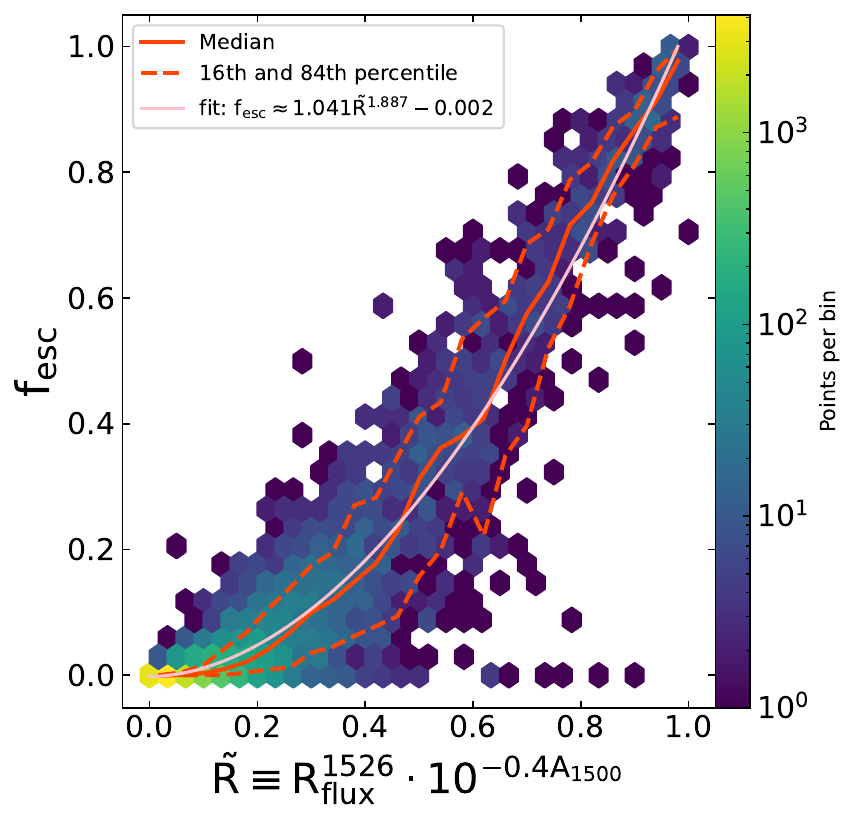}}
  \caption{Comparison of the escape fraction of ionising photons with the product of $\siiilineb$ residual flux and the dust attenuation. The colour scale shows the number of spectra in each hexagonal bin. The solid orange line shows the running median, and the dashed lines show the $\sixteenth$ and $\eightyfourth$ percentiles. The pink line shows the best power function fit to the data. While most of the points lie on the bottom left corner, around $5\%$ of them have $\rm{\tilde{R}}>0.4$ and $\fesc>0.2$, and almost $20\%$ of the galaxies in the simulations contribute to these strongly leaking points.
  }
  \label{fig:rflux_old_paper}
\end{figure}

Following \citetalias{Mauerhofer21} and the picket-fence model \citep[e.g.][]{Gazagnes18, Saldana-Lopez22}, we applied a dust correction factor to the residual flux of absorption lines to predict a more accurate value for $\fesc$. A better measure of the fraction of light that is absorbed by intervening dust and gas is not $\rm{R_{flux}^{1526}}$, but the ratio of the flux at the bottom of the line over the flux of the intrinsic continuum. When we assume, as in the picket-fence model, that galaxies are partially covered by dense optically thick (in dust, $\hzero$ and $\Siplus$) gas, while the rest consists of empty holes, then the escape fraction of ionising photons $\fesc$ is indeed exactly equal to the ratio of the flux at the bottom of LIS absorption lines over the intrinsic continuum flux. Even though galaxies are not true picket fences, we show in Fig.\ref{fig:rflux_old_paper} that we still obtain a much better approximation of $\fesc$ with this method. In this figure, the x-axis represents the residual flux of $\siiilineb$ multiplied by the dust attenuation at 1500 $\angstrom$. A direct comparison of spectra generated without dust would provide a complementary more idealised test, but this is beyond the scope of this work, which is focused on observations.

For comparison, we show the same results in Appendix \ref{sec:1260_plots}, but using $\siiiline$ instead of $\siiilineb$. As we stated above, the former line yields poorer results with our method, with an increase of $42\%$ in the prediction error of $\fesc$.

We performed a fit of the data with a power function, which is shown by the pink line, and accurately fits the median relation (solid orange line). It has the form
\be \label{eq:pred_fesc}
\fesc \approx (1.041 \pm 0.0034) \tilde{R}^{1.887 \pm 0.0087} - (0.002 \pm 0.0004),
\ee
where $\tilde{R}$ is the dust-correct residual flux, computed as $\rm{R_{flux}^{1526}} \cdot 10^{-0.4A_{1500}}$, and $A_{1500}$ is the dust attenuation at $1500 \, \angstrom$ in units of magnitudes. This yields an average error on $\fesc$ of 0.0154 (0.0027 for $\tilde{R}<0.1$ and 0.0459 for $\tilde{R}>0.1$). To further quantify the accuracy of this method, we defined a leaking sightline as having an escape fraction of ionising photons higher than $10\%$. The completeness of this method for detecting leakers, that is, the fraction of leakers that we correctly identify compared to the total number of leakers, is $81.6\%$. The precision, which is the ratio of true positives over the sum of true positives and false positives, is $81.1\%$. 

Finally, we applied the same procedure to predict $\fesclimit$ instead of $\fesc$. We showed in Fig.\ref{fig:fesc_fesc_900} that the two quantities are close to each other, so we do not show a new figure here. The relation we found for the prediction of $\fesclimit$ from the residual flux of $\siiilineb$ and the dust attenuation factor is
\be \label{eq:pred_fesc_900}
\fesclimit \approx (1.018 \pm 0.0037)\tilde{R}^{2.098 \pm 0.0107} - (0.0008 \pm 0.0004).
\ee
For clarity, all the errors discussed here are summarised in $\rm{Table \, \ref{tab:fesc_errors}}$, which also includes the results obtained below for observationally derived dust attenuations (Sections \ref{sec:cigale_lephare_ficus} and Appendix \ref{sec:lephare}). The table shows that the completeness and precision for $\fesclimit$ are similar to that for $\fesc$, and the mean error on $\fesclimit$ is only slightly smaller than the mean error on $\fesc$. In the table, we also add mean absolute errors for probable leakers ($\tilde{R} > 0.1$) and likely non-leakers ($\tilde{R} < 0.1$). The error is naturally larger for probable leakers because the $\fesc$ values are higher. We chose to give the errors not in relative terms because then, predicting an escape fraction of $10^{-2}$ when the true value is $10^{-3}$ would yield a relative error of $900\%$, which is a large error, even though the prediction correctly assessed that the galaxy leaks only a small fraction of ($\leq 1\%$) of ionising photons.

\begin{table}
\caption{Comparison of $f_{\mathrm{esc}}$ prediction accuracy and leaker classification performance using different $A_{1500}$ estimates.}
\centering
\renewcommand{\arraystretch}{1.2}
\setlength{\tabcolsep}{4pt}
\begin{tabular}{lcccc}
\hline\hline
 & \multicolumn{4}{c}{$A_{1500}$ from} \\
\cline{2-5}
 & Simulation & \textsc{Cigale} & \textsc{lephare} & \textsc{ficus} \\
\hline
Mean error on $f_{\mathrm{esc}}$ & 0.0154 & 0.0238 & 0.0215 & 0.0243 \\
Mean error ($\tilde{R} < 0.1$)           & 0.0027 & 0.0030 & 0.0028 & 0.0030 \\
Mean error ($\tilde{R} > 0.1$)           & 0.0459 & 0.0513 & 0.0566 & 0.0634 \\
Completeness                     & 81.6\% & 87.3\% & 80.7\% & 76.6\% \\
Precision                        & 81.1\% & 67.6\% & 78.3\% & 75.9\% \\
\hline
Mean error on $\fesclimit$ & 0.0141 & 0.020 & 0.0186 & 0.0208 \\
Mean error ($\tilde{R} < 0.1$)           & 0.0016 & 0.0018 & 0.0017 & 0.0017 \\
Mean error ($\tilde{R} > 0.1$)           & 0.0440 & 0.0443 & 0.0505 & 0.0558 \\
Completeness ($\fesclimit$)      & 80.8\% & 87.3\% & 79.7\% & 76.0\% \\
Precision ($\fesclimit$)         & 81.6\% & 69.3\% & 79.7\% & 78.8\% \\
\hline
\end{tabular}
\tablefoot{The mean errors are absolute errors, $\langle |\fesc - f_{\rm{esc}}^{\rm{predicted}}| \rangle$. The completeness and precision values refer to the identification of galaxies with $\fesc>0.1$ (leakers). Completeness is the number of true positives divided by the sum of true positives and false negatives, while precision is the number of true positives divided by the sum of true positives and false positives.
}
\label{tab:fesc_errors}
\end{table}

%%%%%%%%%%%%%%%%%%%%%%%%%%%%%%%%%%%%%%%%%%%%%%%%%%%%%%%%%%%%%%%%%%%%%%%%%%%%%%%%%
\section{Discussion} \label{sec:discussion}

Based on our numerous highly resolved simulated galaxies, we have demonstrated that the knowledge of the residual flux of LIS absorption lines, in particular, of $\siiilineb$, may allow us to infer accurate escape fractions of ionising photons provided that we can perfectly correct for dust attenuation.
When we applied this method to actual observations, a few limitations appeared that we discuss in Sections \ref{sec:obs_residual_flux} and \ref{sec:cigale_lephare_ficus}. In Section \ref{sec:comp_obs} we then
apply our predictions to observed data for which the escape fraction was determined independently in order to test our method.

%......................................................................................
\subsection{Measurement of the residual flux} \label{sec:obs_residual_flux}

The residual flux of absorption lines is challenging to measure observationally because the line-spread function and low resolution can smooth absorption profiles and thus artificially increase the residual flux. This has been extensively studied in \cite{Jennings25}, who applied observational effects to simulated spectra from the zoom-in simulation presented in \citetalias{Mauerhofer21}, produced with the same mock observation method as in this paper. They degraded the simulated spectra following three different surveys characteristics, (a) LzLCS G140L \citep{LzLCS_Flury_1}, (b) CLASSY \citep{Berg22}, and (c) VANDELS \citep{Garilli21}, whose resolutions (in terms of the full width at half-maximum $\sigma_{\rm{FWHM}}$) are $300 \ \kms$, $65 \ \kms$, and $461 \ \kms$, respectively.

Their figures 2 and 3 show that the resolution of LzLCS G140L and VANDELS is insufficient for capturing the residual flux of $\siiiline$ correctly. While the equivalent width is correctly inferred on average, the residual flux is mostly overestimated by $\sim 0.3-0.4$ (absolute error), and sometimes up to $\sim 0.8$ or underestimated by $\sim 0.2$. For CLASSY, in contrast, the high resolution allows for a relatively faithful measurement of the residual flux, with errors generally around 0.05. Figure 5 of \cite{Jennings25} also shows that a stacking analysis introduces another error in the measurement of the residual flux.

These errors explain a significant part of the discrepancy that we show in Section \ref{sec:comp_obs}, since we did not apply observational effects when measuring the residual flux from our mock data.

%......................................................................................
\subsection{Observational determination of the dust extinction} \label{sec:cigale_lephare_ficus}

An additional uncertainty in the derivation of $\fesc$ is created by the biases and errors introduced by observational methods for deriving the UV dust attenuation factor. To quantify this, we tested three different codes (Cigale, LePhare, and FiCus) on our mock observations.

\subsubsection{\textsc{Cigale}} \label{sec:cigale}

\begin{figure}
  \resizebox{\hsize}{!}{\includegraphics{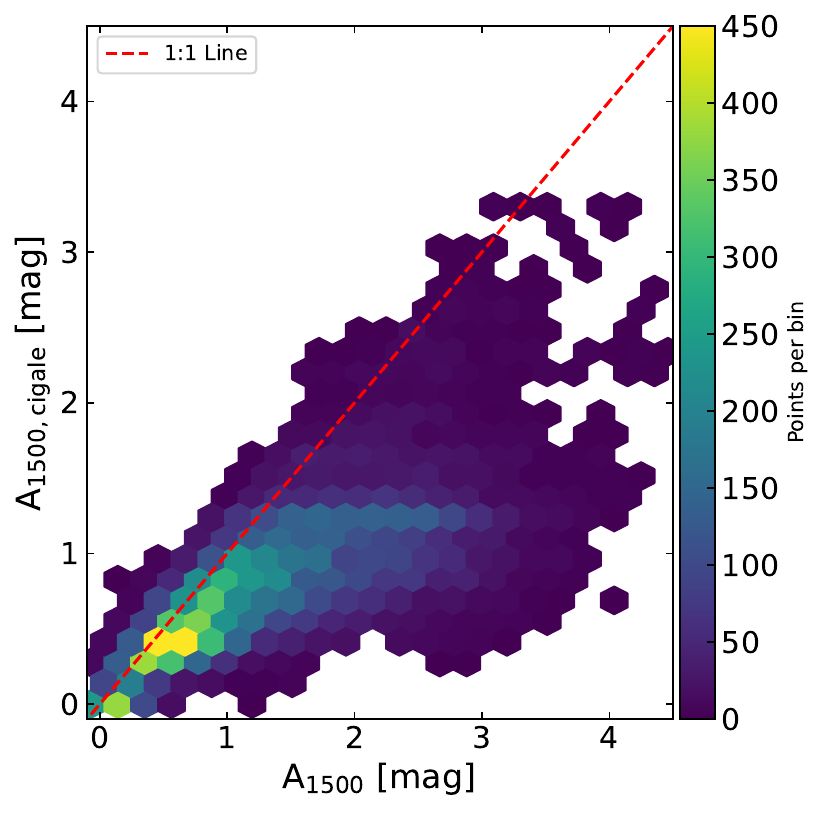}}
  \caption{Dust attenuation at $1500\,\angstrom$ predicted by \textsc{Cigale} compared to the true attenuation (see details of the \textsc{Cigale} implementation in Sect. \ref{sec:cigale}).}
  \label{fig:A1500_cigale}
\end{figure}

In order to reproduce a typical way of deriving the dust attenuation in observations of star-forming galaxies, we used the SED-fitting tool Code Investigating GALaxy Emission \citep[\textsc{Cigale}\footnote{\url{https://cigale.lam.fr/}}: ][]{Cigale}
applied to the 20 NIRCAM filter values provided in \cite{Sphinx20_release} for our simulated galaxies. We assumed a double-exponential shape of the star formation history (SFH), a Chabrier IMF, the BPASS stellar library, and the SMC extinction law. We fixed the galaxy redshifts to their true value, and we allowed for a large number of values for $\rm{E_{B-V}}$. 
We note that for our galaxies at redshifts 4.64 and 5, we removed the bluest filter, NIRCAM F090W, from the fits because \cite{Sphinx20_release} assumes that all photons below the rest-frame $\rm{Lyman}$-$\alpha$ wavelength are absorbed by the IGM. However, at these two lowest {\sc sphinx}$^{20}$ redshifts, this approximation is too strong because a non-negligible fraction of photons below this wavelength can propagate through the IGM. Since the bluest filter overlaps with wavelengths below the rest-frame $\rm{Lyman}$-$\alpha$ at these two redshifts, we did not take the corresponding flux values into account.

We show the resulting predictions of the attenuation $\rm{A_{1500}}$ compared to the true values in Fig.\ref{fig:A1500_cigale}.
For most spectra with $A_{1500}<1$, the \textsc{Cigale} predictions of the attenuation are relatively accurate, although a few of them show a difference of up to one magnitude compared to the true value. For spectra with higher attenuations, \textsc{Cigale} predictions are biased towards lower values: they underestimate the true attenuation. 
To study the reasons for this, we selected 140 galaxies with $A_{1500} \approx 2$ mag and computed the dust optical depth in front of all the stellar particles of these galaxies by tracing rays with {\sc rascas}. The distribution of the dust optical depth shows that galaxies with a high optical depth in front of the brightest star clusters always have their dust attenuation underestimated by \textsc{Cigale} by at least one magnitude. Galaxies with a more homogeneous cover of dust in front of all stars can be either correctly predicted or not. It is understandable that $A_{1500}$ is underestimated by \textsc{Cigale} for galaxies with optically thick dust in front of bright star clusters because these clusters do not contribute to the UV spectra at all (and often not even to the optical spectra), and thus, are missed by the SED fitting. A knowledge of the dust emission in the infrared would help us to better constrain the dust attenuation.

We recomputed the errors of our models (Eq. \ref{eq:pred_fesc}), except with $A_{1500}$ predicted by \textsc{Cigale} instead of the true value from the simulation, and we list the answers in Table \ref{tab:fesc_errors}. We obtain an average error on $\fesc$ of 0.0238 (0.003 for $\tilde{R}<0.1$ and 0.0513 for $\tilde{R}>0.1$). For the detection of leakers (with $\fesc>0.1$), the completeness increases from $81.6\%$ to $87.3\%$, meaning that more true leakers are correctly identified. However, the precision decreases significantly, from $81.1\%$ to $67.6\%$, which means that almost one-third of the galaxies identified as leakers actually have low escape fractions. This is expected because the dust attenuation is underestimated by \textsc{Cigale} on average, and Equation \ref{eq:pred_fesc} shows that $\fesc$ is overestimated. 
While less accurate than when using the real dust attenuation, these predictions of $\fesc$ are still mostly reliable and useful. We found similar results with another SED-fitting code, \textsc{lephare}, which we show in $\rm{Appendix} \ \ref{sec:lephare}$.

\subsubsection{\textsc{FiCUS}} \label{sec:ficus}

\begin{figure}
  \resizebox{\hsize}{!}{\includegraphics{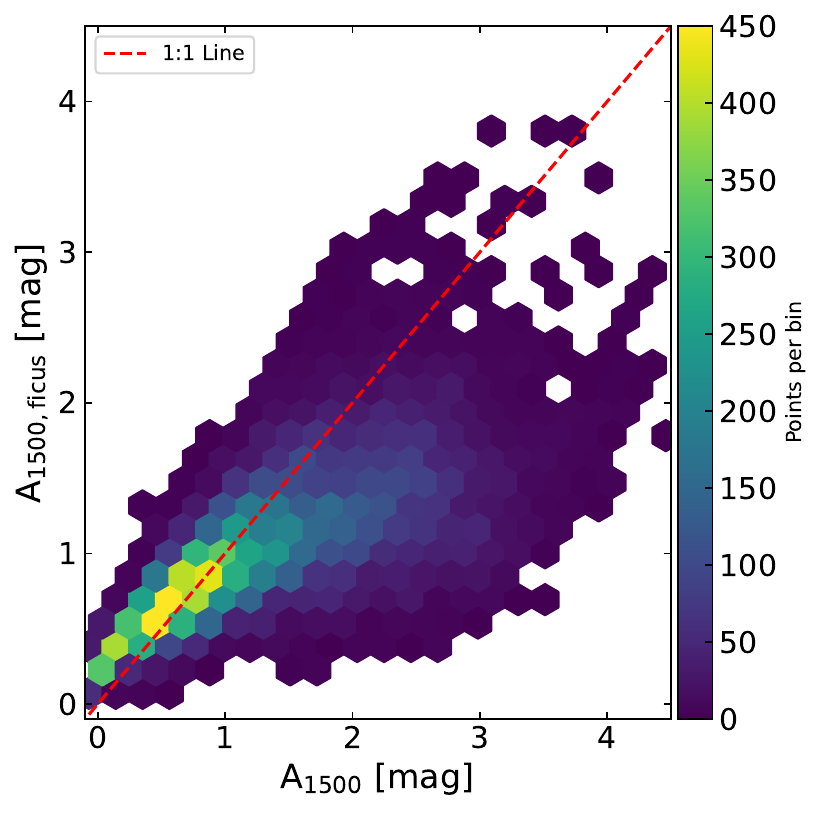}}
  \caption{Dust attenuation predicted by \textsc{ficus} compared to the true attenuation.}
  \label{fig:A1500_ficus}
\end{figure}

Finally, we also tested a different approach, using the spectral fitting tool FItting the stellar Continuum of Uv Spectra \citep[\textsc{ficus}\footnote{\url{https://github.com/asalda/ficus}}: ][]{Saldana-Lopez23}, applied to {\sc sphinx}$^{20}$ continuum spectra from $1200\,\angstrom$ to $2000\,\angstrom$, and assuming the SMC extinction law. These spectra were obtained with {\sc rascas}, including only the interaction of photons with dust, not with metallic ions, since LIS absorption lines are masked by \textsc{ficus} in any case. The fiducial version of \textsc{ficus} uses a combination of \textsc{starburst99} \citep{Leitherer99} and \textsc{bpass} stellar templates, including a model of the nebular continuum, but since we did not model this nebular component, we used a modified version of \textsc{ficus} with the stellar continuum from \textsc{bpass} alone. The predictions of \textsc{ficus} for the dust attenuation are shown in Fig.\ref{fig:A1500_ficus}. The results we found are similar to those for the SED-fitting codes in that high $A_{1500}$ values are underestimated by \textsc{ficus}. Additionally, we found that it slightly overestimates $A_{1500}$ in the low dust attenuation regime. Thus, the mean errors on $\fesc$ are slightly larger than for \textsc{Cigale} and \textsc{lephare}, at 0.0243 (0.0030 for $\tilde{R}<0.1$ and 0.0634 for $\tilde{R}>0.1$; see more details in Table \ref{tab:fesc_errors}).

In summary, all three methods show that they cannot infer the intrinsic UV continuum accurately when $A_{1500} \gtrsim 1 \text{--} 1.5$. This is an expected result because strongly attenuated zones in a galaxy are strongly dominated in the integrated spectrum by low-attenuation zones \citep[e.g.][]{Gazagnes23}. Despite this, the errors they introduce when inferring the escape fraction with our new method are small. This is in part due to the fact that galaxies with strong dust attenuation, for which the inferred $A_{1500}$ are the most inaccurate, usually have almost no escape of ionising photons ($\fesc \approx 0$).

%......................................................................................
\subsection{Applying our method to observations} \label{sec:comp_obs}

Equipped with a simple relation between absorption line depth, dust attenuation, and escape fraction, we applied our predictions to observed data of galaxies with independent estimates of $\fesc$ to test whether the limitations listed above still result in a good agreement. Since only the LyC close to the Lyman limit can be observed, we used our Equation \ref{eq:pred_fesc_900} to predict $\fesclimit$.

\subsubsection{LzLCS} \label{sec:comp_lzlcs}

\begin{figure}
  \resizebox{\hsize}{!}{\includegraphics{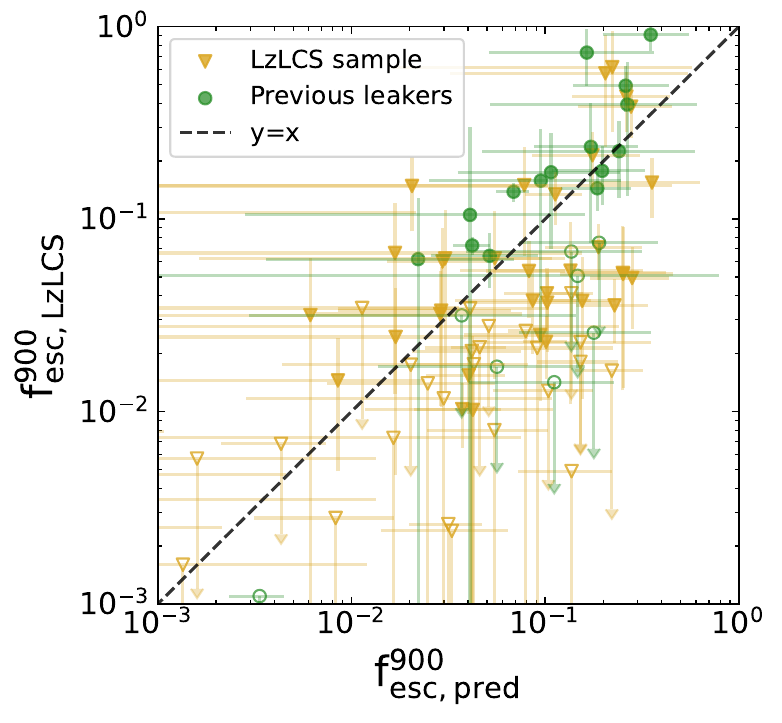}}
  \caption{Comparison of Equation \ref{eq:pred_fesc_900} applied to LzLCS residual fluxes and dust attenuations (x-axis) with $\fesclimit$ estimated in \cite{LzLCS_Flury_1} with the UV SED $\fesc$ method (y-axis). The yellow points correspond to galaxies from the LzLCS survey itself, while the green points correspond to leakers detected earlier \citep{Izotov16a, Izotov16b, Izotov18b, Izotov21, Wang19}. The upper limits are indicated with open symbols and downward arrows.}
  \label{fig:lzlcs_comp}
\end{figure}

The LzLCS sample \citep{LzLCS_Flury_1, LzLCS_Flury_2} contains 66 star-forming galaxies observed with the Cosmic Origins Spectrograph (COS) on the Hubble Space Telescope (HST), including coverage on the LyC. It is an ideal dataset to test our predictions of the escape fraction of ionising photons. Using a public database of LzLCS galaxy properties in addition to 23 leakers compiled from previous studies \citep{Izotov16a, Izotov16b, Izotov18b, Izotov21, Wang19}, we read the UV dust attenuation estimated for each galaxy (assuming the SMC extinction law) as well as the residual flux of absorption lines. Unfortunately, the $\siiilineb$ line is not observed in this sample because it falls outside the wavelength range of HST/COS, so we instead relied on an average residual flux of all LIS lines they detected ($\siiidoublet$, $\siiiline$, $\oilinesiii$, and $\ciiline$), called $\rm{R_{LIS}}$. Assuming the error on dust attenuation and residual flux provided by the LzLCS data follows a Gaussian distribution, we estimated the error on the escape fraction via a Monte Carlo sampling of Equation \ref{eq:pred_fesc_900}. This prediction of $\fesclimit$ can then be compared to the escape fraction inferred from the LyC of the LzLCS sample. 
 \cite{LzLCS_Flury_1} presented three different ways of inferring the escape fractions. We adopted the UV SED $\fesc$, which they determined to be the most reliable. It is the ratio of the measured LyC flux to the intrinsic LyC flux given by the best-fit \textsc{starburst99} templates computed with \textsc{ficus}, and we call it $\rm{f_{esc,LzLCS}^{900}}$. We present the comparison of our prediction of LzLCS escape fractions ($\rm{f_{esc,pred}^{900}}$) versus the measurements of \cite{LzLCS_Flury_1} in Fig.\ref{fig:lzlcs_comp}. The match is good overall, although the error bars are large, and for some galaxies, the prediction does not match the measured $\fesclimit$ even within the errors. For example, the eight galaxies with $\rm{f_{esc,LzLCS}^{900}} > 0.3$ are predicted to have values that are lower by two to four times with our method. Additionally, the vast majority of galaxies with upper limits on $\fesclimit$ in LzLCS (empty symbols in Fig.\ref{fig:lzlcs_comp}) are predicted to have non-negligible escape fractions, above a few percent. 

These discrepancies can be understood from several factors. We showed in Sect. \ref{sec:obs_residual_flux} that the LzLCS survey characteristics tend to degrade absorption lines, leading to inaccurate determinations of the residual flux. In particular, the absorption is spread, artificially increasing the residual flux. This effect would cause our predictions to be higher than the actual values. Additionally, we showed in Sect. \ref{sec:ficus} that the measurements of dust attenuation with observed UV spectra are relatively inaccurate. The dust attenuation determinations for LzLCS galaxies are indeed performed with \textsc{ficus}. Fig.\ref{fig:A1500_ficus} shows that \textsc{ficus} slightly overestimates $A_{1500}$ on average for galaxies with no or very low dust attenuation. From Equation \ref{eq:pred_fesc_900}, this causes our prediction of $\fesclimit$ to underestimate the true escape fraction. This partly explains the mismatch for the eight leakiest galaxies in Fig.\ref{fig:lzlcs_comp}, which indeed have a very low dust attenuation. In contrast, for galaxies with strong dust attenuation, Fig.\ref{fig:A1500_ficus} shows that \textsc{ficus} underestimates $A_{1500}$, which causes Equation \ref{eq:pred_fesc_900} to overestimate $\fesclimit$. This can explain why many LzLCS upper limits are predicted to have non-negligible escape fractions by our method.
An additional reason for the underestimation of $\fesc$ for the eight leakiest galaxies might be that in our simulations, we did not allow for the escape of nebular LyC photons, which \cite{Izotov25} found to be a non-negligible component for several leakers observed with COS.
Another source of discrepancy, as mentioned above, comes from the fact that we were unable to use $\siiilineb$ for our analysis because it is not in the LzLCS dataset. It is unclear how much this affects our predictions, but the residual fluxes of different LIS lines of oxygen, silicon, and carbon have been shown to be relatively similar \citep[e.g.][]{Steidel18, Parker24}. Therefore, we expect that this caveat in our work does not introduce a significant systematic bias, but only contributes slightly to the scatter in the data.
Finally, another source of discrepancy between the two quantities in Fig.\ref{fig:lzlcs_comp} is the fact that our equation for the prediction of $\fesclimit$ is based on the true escape fractions from simulated galaxies, while this quantity is not directly measurable. For a more direct comparison, the methods of \cite{LzLCS_Flury_1} should be applied to infer $\fesclimit$ directly to mock data of our simulated galaxies. This is beyond our scope, in part because not all the relevant mock data have been created yet. 

\subsubsection{CLASSY} \label{sec:comp_classy}

Another sample of galaxies to test our method comes from the CLASSY survey \citep{Berg22, James22}, which consists of 45 low-redshift ($z<0.18$) star-forming galaxies with diverse characteristics in terms of mass, star formation rate, dust attenuation, and so on. With a maximum resolution of $R \sim15'000$ \citep{Berg22}, the residual flux of absorption lines can be measured with higher precision than for LzLCS galaxies ($R\sim 1000$), and the wavelength range covered by CLASSY allowed us to use the $\siiilineb$ line for almost all its galaxies. However, the redshifts of CLASSY galaxies are too low for HST/COS to be able to detect ionising photons. Instead, we followed \cite{Parker26}, who applied numerous indirect methods for determining $\fesclimit$ and applied them to the high-quality CLASSY spectra. We summarise the escape fractions from that study below.
\begin{itemize}
    \item $f_{\rm{esc}}^{Cf}$ uses LIS absorption lines to infer the covering fraction of LIS metals and in turn of $\hzero$ to derive a value of $\fesclimit$. This is based on \cite{Chisholm18, Saldana-Lopez22, Parker24}.
    
    \item $f_{\rm{esc}}^{\beta}$ uses the trend between the $\beta$ slope of the UV continuum and $\fesclimit$, as found by \cite{Chisholm22}.

    \item $f_{\rm{esc}}^{V_{\rm{sep}}}$ uses the peak separation of the $\rm{Lyman}$-$\alpha$ line, when detected, which was shown to trace $\fesclimit$ by numerous studies \cite[e.g.][]{Verhamme17, Izotov18b, LzLCS_Flury_2}, although the trend was found to not work at higher redshifts \citep[e.g.][]{Kerutt24}.

    \item $f_{\rm{esc}}^{\rm{AFT}}$ uses multivariate statistical analysis performed in \cite{Jaskot24a, Jaskot24b}, based on many properties such as stellar mass, $\rm{E(B-V)}_*$, $\rm{E(B-V)}_{\rm{neb}}$, SFR, the star formation rate surface density, M$_{1500}$, $\beta_{\rm{UV}}$, the ionisation ratio O$_{32}$, and the equivalent width of the H$_{\beta}$ emission line.

    \item $f_{\rm{esc}}^{\rm{sim}}$ uses results from \cite{Gazagnes23}, who fitted mock $\ciiline$ and $\siiiline$ absorption lines to CLASSY galaxies. These mock spectra were obtained from the cosmological zoom-in simulation presented in \citetalias{Mauerhofer21}, using the same method as presented in Sect. \ref{sec:abs_lines}. Processing the simulated galaxy at different times and from many viewing angles, this provides a catalogue of 22500 mock spectra for both lines. For 38 of the 45 CLASSY galaxies, an excellent match is found in this catalogue by fitting both lines simultaneously. For these galaxies, $f_{\rm{esc}}^{\rm{sim}}$ is defined as the escape fraction of the simulated galaxy at the time and viewing angle of the best-match spectrum.

    \item $f_{\rm{esc}}^{\rm{O}_{32}}$ uses the emission line ratio O$_{32}$ = [O$\textsc{iii}$] $\lambda5007$ / [O$\textsc{ii}$] $\lambda3727$, which was shown to correlate with $\fesclimit$ in some studies \citep[e.g.][]{Izotov16a, Izotov18b, Nakajima20}.

    \item $\langle f_{\rm{esc}}^{\rm{LyC}} \rangle$ is the median escape fraction from \cite{Parker26}, obtained by taking the median of the six previous $\fesclimit$ estimates. The uncertainties for this median are defined as the 16th and 84th percentiles from 300 Monte Carlo variations of the individual escape fraction predictions based on their uncertainties.
    
\end{itemize}

Not all these six indirect tracers of $\fesclimit$ are fully independent becaue both $f_{\rm{esc}}^{\rm{AFT}}$ and $f_{\rm{esc}}^{\rm{O}_{32}}$ use the ratio O$_{32}$, $\beta_{\rm{UV}}$ is used in $f_{\rm{esc}}^{\rm{AFT}}$ and $f_{\rm{esc}}^{\beta}$, and LIS absorption lines are used in $f_{\rm{esc}}^{Cf}$ and $f_{\rm{esc}}^{\rm{sim}}$. The correlations between the different methods, as well as their biases as a function of dust attenuation or neutral gas covering fraction, are analysed in detail in \cite{Parker26}.

\begin{figure*}
  \resizebox{\hsize}{!}{\includegraphics{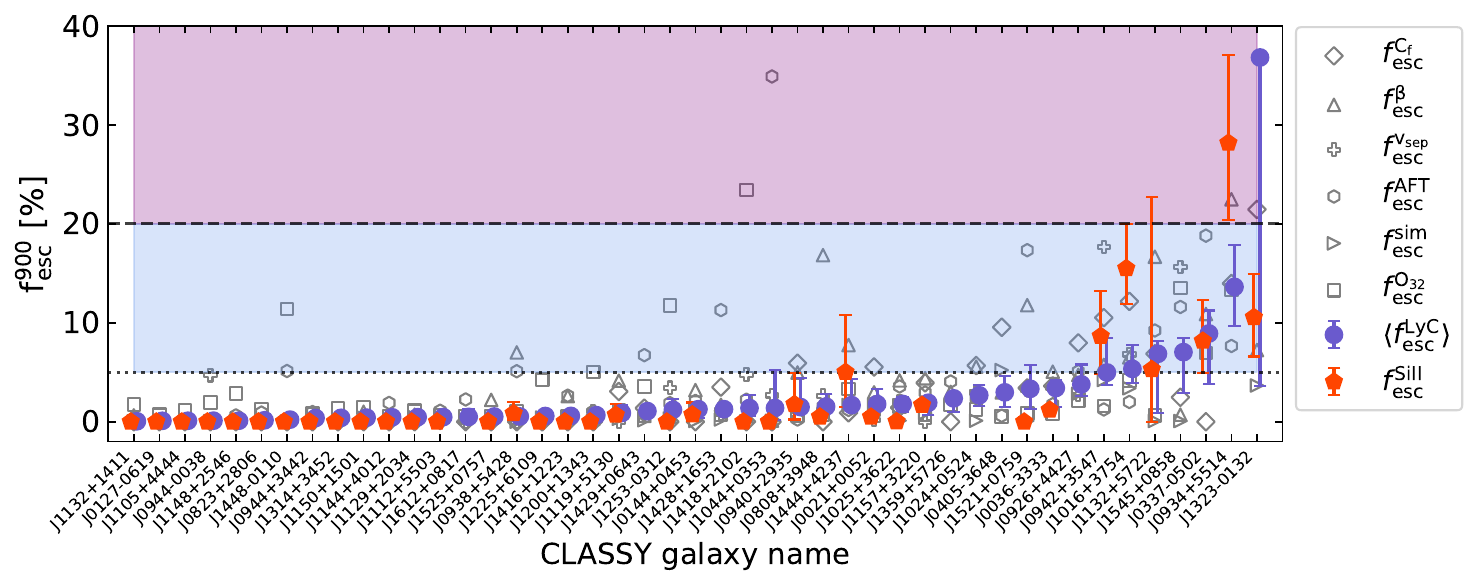}}
  \caption{Comparison of our predictions of $\fesclimit$ from Equation \ref{eq:pred_fesc_900} with escape fractions inferred in Parker et al. (in prep) for all CLASSY galaxies, listed on the x-axis. The open symbols represent the six methods with which $\fesclimit$ was inferred from observations and which are detailed in Sect. \ref{sec:comp_classy}. The purple dots show the median of the six measurements, and the red pentagons show our predictions. The shaded blue region represents weak leakers, as defined in \cite{Parker26}, and the shaded dark pink region represents strong leakers.}
  \label{fig:classy_comp}
\end{figure*}

Our predictions of $\fesclimit$ for CLASSY galaxies are labelled $f_{\rm{esc}}^{\rm{Si\textsc{ii}}}$, and, along with error bars, were computed as in Sect. \ref{sec:comp_lzlcs}, except that we used the $\siiilineb$ line. A few CLASSY galaxies do not include this line, and we therefore omitted them. Comparisons with the escape fractions of \cite{Parker26} are shown in Fig.\ref{fig:classy_comp}, in order of increasing $\langle f_{\rm{esc}}^{\rm{LyC}} \rangle$. Because our predictions are ill-suited for distinguishing escape fractions below one percent, we used a linear scale for the y-axis. For clarity, our main comparison was made with the median $\langle f_{\rm{esc}}^{\rm{LyC}} \rangle$ (purple points), while the points from the six methods presented above are shown in grey for completeness. The figure shows that our predictions are close to the median, $\langle f_{\rm{esc}}^{\rm{LyC}} \rangle$. For most galaxies below the weak leaker threshold of $5\%$, our predictions match the median $\fesclimit$ within the error bars. A slight discrepancy arises only for J1521+0759. For one weak leaker, J1016+3754, our method predicts larger escape fractions, around $15\%$, compared to $5\%$ for the median of \cite{Parker26}. Our prediction is compatible with $f_{\rm{esc}}^{Cf}$, which might indicate that LIS absorption lines in this faint ($\rm{M}_{1500}\sim-14$) galaxy overpredict the true escape fraction. This might be due to its low metallicity ($12+\log(\rm{O/H}) \sim 7.5$), which can result in weak LIS absorption lines, while $\hzero$ still absorbs many ionising photons. Our method also overestimates the escape fraction of J0934+5514, which is an even fainter and more metal-poor galaxy. Overall, the global compatibility of our predictions with the many methods presented in \cite{Parker26} is encouraging for the use of LIS residual flux and dust attenuation to predict the escape fraction of ionising photons at the epoch of reionisation.

%%%%%%%%%%%%%%%%%%%%%%%%%%%%%%%%%%%%%%%%%%%%%%%%%%%%%%%%%%%%%%%%%%%%%%%%%%%%%%%%%
\section{Summary and conclusions} \label{sec:summary}

We have extended the {\sc sphinx}$^{20}$ data release with high-resolution synthetic LIS absorption lines, modelled with full radiative transfer and including detailed physics such as resonant scattering, dust interactions, turbulence, and absorption from the fine-structure level. Using this dataset, we analysed correlations between $\siiilineb$ absorption properties, dust attenuation, and the escape fraction of ionising photons (\(\fesc\)). We also tested how well our results apply to observational data and assessed common observational techniques for inferring dust attenuation.
Our main conclusions are listed below.

\begin{itemize}
  \item We found a tight correlation between the dust-corrected residual flux of the $\siiilineb$ line and the escape fraction of ionising photons, which we expressed as a simple empirical power law (Eq.~\ref{eq:pred_fesc}). This relation predicts \(\fesc\) with an average error below 0.02.

  \item The predictive power of this method is strongest when using high-resolution spectra and accurate estimates of the UV dust attenuation. When using true attenuation values from the simulation, our method achieves 82\% precision and completeness in identifying leakers (\(\fesc > 10\%\)).

  \item We showed that several commonly used methods to infer $A_{1500}$, namely \textsc{Cigale}, \textsc{lephare}, and \textsc{ficus}, often underestimate the true attenuation, especially in dusty galaxies where bright star-forming regions are heavily obscured. This leads to systematic overestimates of \(\fesc\) when applying our formula. However, since most leakers have relatively low dust attenuation, using observationally inferred values of the dust attenuation usually does not introduce strong errors for them.

  \item Application of our method to the LzLCS survey yielded mixed results, with significant scatter and poor agreement for several galaxies, while the escape fraction of others was correctly predicted. We attribute the discrepancies to the limited spectral resolution of the HST/COS G140L grating, the lack of the $\siiilineb$ line, and uncertain dust estimates from the SED-fitting codes.

  \item In contrast, when we applied the same method to CLASSY galaxies, where both $\siiilineb$ and high-resolution UV spectra are available, the results agree well on average with the median \(\fesclimit\) values inferred from six independent (indirect) observational techniques.

  \item The new dataset of LIS absorption lines, combined with ionising escape fractions and galaxy properties, provides a valuable public resource for testing ISM diagnostics and spectrum-fitting techniques. The data are made publicly available (see the section data availability).

  \item Overall, our work supports the use of LIS absorption lines, in particular $\siiilineb$, as observational tracers of \(\fesc\) at the epoch of reionisation, but it highlights that their effectiveness critically depends on our ability to reliably measure the line depth and dust attenuation.
\end{itemize}

Despite the robustness of our results, several caveats remain. First, while our predictive formula performs well within the simulation, it is calibrated on a specific model of galaxy formation, feedback, and dust attenuation \citep{Sphinx20}. Applying it to observations assumes that similar physical conditions hold. Second, our modelling of the silicon ion density assumes solar abundance ratios scaled to local metallicity, along with equilibrium ionisation fractions computed using a fixed radiation field. Both assumptions may introduce uncertainties, particularly in high-redshift and low-density environments. Lastly, the simulation adopts an SMC-like extinction curve and a fixed dust-to-metal scaling, which may not capture the full diversity of dust physics in real galaxies. Processes such as gas–dust decoupling, dust destruction by sputtering, and shocks may lead to different dust opacities and different star-dust geometries compared to our current work. This might alter the quantitative relation between LIS residual flux and $\fesc$, for example by introducing additional scatter or modifying the normalisation of the inferred trends. However, the magnitude and direction of these effects are currently uncertain, and further developments in dust modelling are necessary before these effects can be analysed.

The approach we presented can be expanded in several directions. A logical next step is to apply this method to larger samples of high-resolution UV spectra, especially with upcoming surveys targeting star-forming galaxies at higher redshifts. Our data might also serve as training sets for machine-learning models designed to infer \(\fesc\) directly from spectra. Additionally, extending this analysis to include emission line diagnostics and their connection to the escape of ionising radiation \citep[e.g.][]{Choustikov24} would help us to unify the absorption and emission line approaches to reionisation-era galaxy studies. Finally, new-generation simulations such as MEGATRON \citep{MEGATRON} remove many of the limitations cited above and might create robuster predictions of the escape fraction of ionising photons.

\section*{Data availability} \label{sec:test}
The spectra, absorption line measurements, and escape fraction values used in this work are publicly available here: \url{https://doi.org/10.5281/zenodo.17723577}.

\begin{acknowledgements}

VM acknowledges support from the NWO grant 016.VIDI.189.162 ('ODIN'). T.K. is supported by the National Research Foundation of Korea (RS-2022-NR070872 and RS-2025-00516961) and also by the Yonsei Fellowship, funded by Lee Youn Jae. The authors thank Harley Katz and Annalisa De Cia for helpful discussions and valuable input.

\end{acknowledgements}

\bibliography{references}

@ARTICLE{Parker26,
       author = {{Parker}, Kaelee S. and {Berg}, Danielle A. and {Chisholm}, John and {Gazagnes}, Simon and {Flury}, Sophia R. and {Carr}, Cody and {Huberty}, Mason and {Jaskot}, Anne E. and {Hayes}, Matthew J. and {Saldana-Lopez}, Alberto and {Hernandez}, Svea and {Nanayakkara}, Themiya and {James}, Bethan L. and {Arellano-C{\'o}rdova}, Karla Z. and {Strom}, Allison L. and {Senchyna}, Peter and {Mingozzi}, Matilde and {Heckman}, Timothy and {Xu}, Xinfeng and {Henry}, Alaina and {Amor{\'\i}n}, Ricardo O. and {Mauerhofer}, Valentin and {Martin}, Crystal L. and {Erb}, Dawn K. and {Skillman}, Evan D. and {Rubin}, Kate H.~R. and {Trevino}, John and {Leitherer}, Claus},
        title = "{CLASSY. XIII. Cutting through the Clouds{\textemdash}Comparing Indirect Tracers of Ionizing Photon Escape}",
      journal = {\apj},
         year = 2026,
        month = jan,
       volume = {997},
       number = {1},
          eid = {98},
        pages = {98},
          doi = {10.3847/1538-4357/ae22d9},
archivePrefix = {arXiv},
       eprint = {2511.15869},
 primaryClass = {astro-ph.GA},
       adsurl = {https://ui.adsabs.harvard.edu/abs/2026ApJ...997...98P}
}

@ARTICLE{Izotov25,
       author = {{Izotov}, Y.~I. and {Schaerer}, D. and {Worseck}, G. and {Guseva}, N.~G. and {Verhamme}, A. and {Simmonds}, C. and {Chisholm}, J.},
        title = "{A great diversity of spectral shapes in the ionising spectra of z {\ensuremath{\sim}} 0.6─1 galaxies revealed by HST/COS and possible detection of nebular LyC emission}",
      journal = {\aap},
         year = 2025,
        month = nov,
       volume = {704},
          eid = {A19},
        pages = {A19},
          doi = {10.1051/0004-6361/202556004},
archivePrefix = {arXiv},
       eprint = {2510.22152},
 primaryClass = {astro-ph.GA},
       adsurl = {https://ui.adsabs.harvard.edu/abs/2025A&A...704A..19I}
}

@ARTICLE{MEGATRON,
       author = {{Katz}, Harley and {Rey}, Martin P. and {Cadiou}, Corentin and {Kimm}, Taysun and {Agertz}, Oscar},
        title = "{The Impact of Star Formation and Feedback Recipes on the Stellar Mass and Interstellar Medium of High-Redshift Galaxies}",
      journal = {The Open Journal of Astrophysics},
         year = 2026,
        month = feb,
       volume = {9},
        pages = {56097},
          doi = {10.33232/001c.156097},
archivePrefix = {arXiv},
       eprint = {2411.07282},
 primaryClass = {astro-ph.GA},
       adsurl = {https://ui.adsabs.harvard.edu/abs/2026OJAp....956097K}
}

@ARTICLE{Wise19,
       author = {{Wise}, John H.},
        title = "{Cosmic reionisation}",
      journal = {Contemporary Physics},
         year = 2019,
        month = apr,
       volume = {60},
       number = {2},
        pages = {145-163},
          doi = {10.1080/00107514.2019.1631548},
archivePrefix = {arXiv},
       eprint = {1907.06653},
 primaryClass = {astro-ph.CO},
       adsurl = {https://ui.adsabs.harvard.edu/abs/2019ConPh..60..145W}
}

@ARTICLE{Rivera-Thorsen22,
       author = {{Rivera-Thorsen}, T.~E. and {Hayes}, M. and {Melinder}, J.},
        title = "{A bottom-up search for Lyman-continuum leakage in the Hubble Ultra Deep Field}",
      journal = {\aap},
         year = 2022,
        month = oct,
       volume = {666},
          eid = {A145},
        pages = {A145},
          doi = {10.1051/0004-6361/202243678},
archivePrefix = {arXiv},
       eprint = {2206.10799},
 primaryClass = {astro-ph.GA},
       adsurl = {https://ui.adsabs.harvard.edu/abs/2022A&A...666A.145R}
}

@ARTICLE{Marques-Chaves22,
       author = {{Marques-Chaves}, R. and {Schaerer}, D. and {Amor{\'\i}n}, R.~O. and {Atek}, H. and {Borthakur}, S. and {Chisholm}, J. and {Fern{\'a}ndez}, V. and {Flury}, S.~R. and {Giavalisco}, M. and {Grazian}, A. and {Hayes}, M.~J. and {Heckman}, T.~M. and {Henry}, A. and {Izotov}, Y.~I. and {Jaskot}, A.~E. and {Ji}, Z. and {McCandliss}, S.~R. and {Oey}, M.~S. and {{\"O}stlin}, G. and {Ravindranath}, S. and {Rutkowski}, M.~J. and {Saldana-Lopez}, A. and {Teplitz}, H. and {Thuan}, T.~X. and {Verhamme}, A. and {Wang}, B. and {Worseck}, G. and {Xu}, X.},
        title = "{No correlation of the Lyman continuum escape fraction with spectral hardness}",
      journal = {\aap},
         year = 2022,
        month = jul,
       volume = {663},
          eid = {L1},
        pages = {L1},
          doi = {10.1051/0004-6361/202243598},
archivePrefix = {arXiv},
       eprint = {2205.05567},
 primaryClass = {astro-ph.GA},
       adsurl = {https://ui.adsabs.harvard.edu/abs/2022A&A...663L...1M}
}

@ARTICLE{Saxena22,
       author = {{Saxena}, A. and {Pentericci}, L. and {Ellis}, R.~S. and {Guaita}, L. and {Calabr{\`o}}, A. and {Schaerer}, D. and {Vanzella}, E. and {Amor{\'\i}n}, R. and {Bolzonella}, M. and {Castellano}, M. and {Fontanot}, F. and {Hathi}, N.~P. and {Hibon}, P. and {Llerena}, M. and {Mannucci}, F. and {Saldana-Lopez}, A. and {Talia}, M. and {Zamorani}, G.},
        title = "{No strong dependence of Lyman continuum leakage on physical properties of star-forming galaxies at {\ensuremath{\lesssim}} z {\ensuremath{\lesssim}} 3.5}",
      journal = {\mnras},
         year = 2022,
        month = mar,
       volume = {511},
       number = {1},
        pages = {120-138},
          doi = {10.1093/mnras/stab3728},
archivePrefix = {arXiv},
       eprint = {2109.03662},
 primaryClass = {astro-ph.GA},
       adsurl = {https://ui.adsabs.harvard.edu/abs/2022MNRAS.511..120S}
}

@ARTICLE{Vanzella18,
       author = {{Vanzella}, E. and {Nonino}, M. and {Cupani}, G. and {Castellano}, M. and {Sani}, E. and {Mignoli}, M. and {Calura}, F. and {Meneghetti}, M. and {Gilli}, R. and {Comastri}, A. and {Mercurio}, A. and {Caminha}, G.~B. and {Caputi}, K. and {Rosati}, P. and {Grillo}, C. and {Cristiani}, S. and {Balestra}, I. and {Fontana}, A. and {Giavalisco}, M.},
        title = "{Direct Lyman continuum and Ly {\ensuremath{\alpha}} escape observed at redshift 4}",
      journal = {\mnras},
         year = 2018,
        month = may,
       volume = {476},
       number = {1},
        pages = {L15-L19},
          doi = {10.1093/mnrasl/sly023},
archivePrefix = {arXiv},
       eprint = {1712.07661},
 primaryClass = {astro-ph.GA},
       adsurl = {https://ui.adsabs.harvard.edu/abs/2018MNRAS.476L..15V}
}

@ARTICLE{Vanzella15,
       author = {{Vanzella}, E. and {de Barros}, S. and {Castellano}, M. and {Grazian}, A. and {Inoue}, A.~K. and {Schaerer}, D. and {Guaita}, L. and {Zamorani}, G. and {Giavalisco}, M. and {Siana}, B. and {Pentericci}, L. and {Giallongo}, E. and {Fontana}, A. and {Vignali}, C.},
        title = "{Peering through the holes: the far-UV color of star-forming galaxies at z \raisebox{-0.5ex}\textasciitilde 3-4 and the escaping fraction of ionizing radiation}",
      journal = {\aap},
         year = 2015,
        month = apr,
       volume = {576},
          eid = {A116},
        pages = {A116},
          doi = {10.1051/0004-6361/201525651},
archivePrefix = {arXiv},
       eprint = {1502.04708},
 primaryClass = {astro-ph.GA},
       adsurl = {https://ui.adsabs.harvard.edu/abs/2015A&A...576A.116V}
}

@ARTICLE{Vanzella10,
       author = {{Vanzella}, E. and {Siana}, B. and {Cristiani}, S. and {Nonino}, M.},
        title = "{Contamination on Lyman continuum emission at z >\raisebox{-0.5ex}\textasciitilde 3: implication on the ionizing radiation evolution}",
      journal = {\mnras},
         year = 2010,
        month = jun,
       volume = {404},
       number = {4},
        pages = {1672-1678},
          doi = {10.1111/j.1365-2966.2010.16408.x},
archivePrefix = {arXiv},
       eprint = {1001.3891},
 primaryClass = {astro-ph.CO},
       adsurl = {https://ui.adsabs.harvard.edu/abs/2010MNRAS.404.1672V}
}

@ARTICLE{Xu23_classy,
       author = {{Xu}, Xinfeng and {Heckman}, Timothy and {Henry}, Alaina and {Berg}, Danielle A. and {Chisholm}, John and {James}, Bethan L. and {Martin}, Crystal L. and {Stark}, Daniel P. and {Hayes}, Matthew and {Arellano-C{\'o}rdova}, Karla Z. and {Carr}, Cody and {Huberty}, Mason and {Mingozzi}, Matilde and {Scarlata}, Claudia and {Sugahara}, Yuma},
        title = "{CLASSY. VI. The Density, Structure, and Size of Absorption-line Outflows in Starburst Galaxies}",
      journal = {\apj},
         year = 2023,
        month = may,
       volume = {948},
       number = {1},
          eid = {28},
        pages = {28},
          doi = {10.3847/1538-4357/acbf46},
archivePrefix = {arXiv},
       eprint = {2301.11498},
 primaryClass = {astro-ph.GA},
       adsurl = {https://ui.adsabs.harvard.edu/abs/2023ApJ...948...28X}
}

@ARTICLE{Garel24,
       author = {{Garel}, T. and {Michel-Dansac}, L. and {Verhamme}, A. and {Mauerhofer}, V. and {Katz}, H. and {Blaizot}, J. and {Leclercq}, F. and {Salvignol}, G.},
        title = "{A public grid of radiative transfer simulations for Ly{\ensuremath{\alpha}} and metal lines in idealised galactic outflows}",
      journal = {\aap},
         year = 2024,
        month = nov,
       volume = {691},
          eid = {A213},
        pages = {A213},
          doi = {10.1051/0004-6361/202450654},
archivePrefix = {arXiv},
       eprint = {2408.03605},
 primaryClass = {astro-ph.GA},
       adsurl = {https://ui.adsabs.harvard.edu/abs/2024A&A...691A.213G}
}

@ARTICLE{Kerutt24,
       author = {{Kerutt}, J. and {Oesch}, P.~A. and {Wisotzki}, L. and {Verhamme}, A. and {Atek}, H. and {Herenz}, E.~C. and {Illingworth}, G.~D. and {Kusakabe}, H. and {Matthee}, J. and {Mauerhofer}, V. and {Montes}, M. and {Naidu}, R.~P. and {Nelson}, E. and {Reddy}, N. and {Schaye}, J. and {Simmonds}, C. and {Urrutia}, T. and {Vitte}, E.},
        title = "{Lyman continuum leaker candidates at z {\ensuremath{\sim}} 3-4 in the HDUV based on a spectroscopic sample of MUSE LAEs}",
      journal = {\aap},
         year = 2024,
        month = apr,
       volume = {684},
          eid = {A42},
        pages = {A42},
          doi = {10.1051/0004-6361/202346656},
archivePrefix = {arXiv},
       eprint = {2312.08791},
 primaryClass = {astro-ph.GA},
       adsurl = {https://ui.adsabs.harvard.edu/abs/2024A&A...684A..42K}
}

@ARTICLE{Nakajima20,
       author = {{Nakajima}, Kimihiko and {Ellis}, Richard S. and {Robertson}, Brant E. and {Tang}, Mengtao and {Stark}, Daniel P.},
        title = "{The Lyman Continuum Escape Survey. II. Ionizing Radiation as a Function of the [O III]/[O II] Line Ratio}",
      journal = {\apj},
         year = 2020,
        month = feb,
       volume = {889},
       number = {2},
          eid = {161},
        pages = {161},
          doi = {10.3847/1538-4357/ab6604},
archivePrefix = {arXiv},
       eprint = {1909.07396},
 primaryClass = {astro-ph.GA},
       adsurl = {https://ui.adsabs.harvard.edu/abs/2020ApJ...889..161N}
}

@ARTICLE{Jaskot24b,
       author = {{Jaskot}, Anne E. and {Silveyra}, Anneliese C. and {Plantinga}, Anna and {Flury}, Sophia R. and {Hayes}, Matthew and {Chisholm}, John and {Heckman}, Timothy and {Pentericci}, Laura and {Schaerer}, Daniel and {Trebitsch}, Maxime and {Verhamme}, Anne and {Carr}, Cody and {Ferguson}, Henry C. and {Ji}, Zhiyuan and {Giavalisco}, Mauro and {Henry}, Alaina and {Marques-Chaves}, Rui and {{\"O}stlin}, G{\"o}ran and {Saldana-Lopez}, Alberto and {Scarlata}, Claudia and {Worseck}, G{\'a}bor and {Xu}, Xinfeng},
        title = "{Multivariate Predictors of Lyman Continuum Escape. II. Predicting Lyman Continuum Escape Fractions for High-redshift Galaxies}",
      journal = {\apj},
         year = 2024,
        month = oct,
       volume = {973},
       number = {2},
          eid = {111},
        pages = {111},
          doi = {10.3847/1538-4357/ad5557},
archivePrefix = {arXiv},
       eprint = {2406.10179},
 primaryClass = {astro-ph.GA},
       adsurl = {https://ui.adsabs.harvard.edu/abs/2024ApJ...973..111J}
}

@ARTICLE{Jaskot24a,
       author = {{Jaskot}, Anne E. and {Silveyra}, Anneliese C. and {Plantinga}, Anna and {Flury}, Sophia R. and {Hayes}, Matthew and {Chisholm}, John and {Heckman}, Timothy and {Pentericci}, Laura and {Schaerer}, Daniel and {Trebitsch}, Maxime and {Verhamme}, Anne and {Carr}, Cody and {Ferguson}, Henry C. and {Ji}, Zhiyuan and {Giavalisco}, Mauro and {Henry}, Alaina and {Marques-Chaves}, Rui and {{\"O}stlin}, G{\"o}ran and {Saldana-Lopez}, Alberto and {Scarlata}, Claudia and {Worseck}, G{\'a}bor and {Xu}, Xinfeng},
        title = "{Multivariate Predictors of Lyman Continuum Escape. I. A Survival Analysis of the Low-redshift Lyman Continuum Survey}",
      journal = {\apj},
         year = 2024,
        month = sep,
       volume = {972},
       number = {1},
          eid = {92},
        pages = {92},
          doi = {10.3847/1538-4357/ad58b9},
archivePrefix = {arXiv},
       eprint = {2406.10171},
 primaryClass = {astro-ph.GA},
       adsurl = {https://ui.adsabs.harvard.edu/abs/2024ApJ...972...92J}
}

@ARTICLE{Parker24,
       author = {{Parker}, Kaelee S. and {Berg}, Danielle A. and {Gazagnes}, Simon and {Chisholm}, John and {James}, Bethan L. and {Hayes}, Matthew and {Heckman}, Timothy and {Henry}, Alaina and {Berg}, Michelle A. and {Arellano-C{\'o}rdova}, Karla Z. and {Xu}, Xinfeng and {Erb}, Dawn K. and {Martin}, Crystal L. and {Hu}, Weida and {Skillman}, Evan D. and {McQuinn}, Kristen B.~W. and {Chen}, Zuyi and {Stark}, Dan P.},
        title = "{CLASSY. XI. Tracing Neutral Gas Properties Using UV Absorption Lines and 21 cm Observations}",
      journal = {\apj},
         year = 2024,
        month = dec,
       volume = {977},
       number = {1},
          eid = {104},
        pages = {104},
          doi = {10.3847/1538-4357/ad87cd},
archivePrefix = {arXiv},
       eprint = {2410.00236},
 primaryClass = {astro-ph.GA},
       adsurl = {https://ui.adsabs.harvard.edu/abs/2024ApJ...977..104P}
}

@ARTICLE{James22,
       author = {{James}, Bethan L. and {Berg}, Danielle A. and {King}, Teagan and {Sahnow}, David J. and {Mingozzi}, Matilde and {Chisholm}, John and {Heckman}, Timothy and {Martin}, Crystal L. and {Stark}, Dan P. and {Aloisi}, Alessandra and {Amor{\'\i}n}, Ricardo O. and {Arellano-C{\'o}rdova}, Karla Z. and {Bayliss}, Matthew and {Bordoloi}, Rongmon and {Brinchmann}, Jarle and {Charlot}, St{\'e}phane and {Chen}, Zuyi and {Chevallard}, Jacopo and {Clark}, Ilyse and {Erb}, Dawn K. and {Feltre}, Anna and {Hayes}, Matthew and {Henry}, Alaina and {Hernandez}, Svea and {Jaskot}, Anne and {Kewley}, Lisa J. and {Kumari}, Nimisha and {Leitherer}, Claus and {Llerena}, Mario and {Maseda}, Michael and {Nanayakkara}, Themiya and {Ouchi}, Masami and {Plat}, Adele and {Pogge}, Richard W. and {Ravindranath}, Swara and {Rigby}, Jane R. and {Scarlata}, Claudia and {Senchyna}, Peter and {Skillman}, Evan D. and {Steidel}, Charles C. and {Strom}, Allison L. and {Sugahara}, Yuma and {Wilkins}, Stephen M. and {Wofford}, Aida and {Xu}, Xinfeng and {Classy Team}},
        title = "{CLASSY. II. A Technical Overview of the COS Legacy Archive Spectroscopic Survey}",
      journal = {\apjs},
         year = 2022,
        month = oct,
       volume = {262},
       number = {2},
          eid = {37},
        pages = {37},
          doi = {10.3847/1538-4365/ac8008},
archivePrefix = {arXiv},
       eprint = {2206.01224},
 primaryClass = {astro-ph.GA},
       adsurl = {https://ui.adsabs.harvard.edu/abs/2022ApJS..262...37J}
}

@ARTICLE{Asplund21,
       author = {{Asplund}, M. and {Amarsi}, A.~M. and {Grevesse}, N.},
        title = "{The chemical make-up of the Sun: A 2020 vision}",
      journal = {\aap},
         year = 2021,
        month = sep,
       volume = {653},
          eid = {A141},
        pages = {A141},
          doi = {10.1051/0004-6361/202140445},
archivePrefix = {arXiv},
       eprint = {2105.01661},
 primaryClass = {astro-ph.SR},
       adsurl = {https://ui.adsabs.harvard.edu/abs/2021A&A...653A.141A}
}

@ARTICLE{James18,
       author = {{James}, B. and {Aloisi}, A.},
        title = "{Tackling the Saturation of Oxygen: The Use of Phosphorus and Sulfur as Proxies within the Neutral Interstellar Medium of Star-forming Galaxies}",
      journal = {\apj},
         year = 2018,
        month = feb,
       volume = {853},
       number = {2},
          eid = {124},
        pages = {124},
          doi = {10.3847/1538-4357/aa9ffb},
       adsurl = {https://ui.adsabs.harvard.edu/abs/2018ApJ...853..124J}
}

@ARTICLE{James14,
       author = {{James}, B.~L. and {Aloisi}, A. and {Heckman}, T. and {Sohn}, S.~T. and {Wolfe}, M.~A.},
        title = "{Investigating Nearby Star-forming Galaxies in the Ultraviolet with HST/COS Spectroscopy. I. Spectral Analysis and Interstellar Abundance Determinations}",
      journal = {\apj},
         year = 2014,
        month = nov,
       volume = {795},
       number = {2},
          eid = {109},
        pages = {109},
          doi = {10.1088/0004-637X/795/2/109},
archivePrefix = {arXiv},
       eprint = {1408.4420},
 primaryClass = {astro-ph.GA},
       adsurl = {https://ui.adsabs.harvard.edu/abs/2014ApJ...795..109J}
}

@ARTICLE{Wang19,
       author = {{Wang}, Bingjie and {Heckman}, Timothy M. and {Leitherer}, Claus and {Alexandroff}, Rachel and {Borthakur}, Sanchayeeta and {Overzier}, Roderik A.},
        title = "{A New Technique for Finding Galaxies Leaking Lyman-continuum Radiation: [S II]-deficiency}",
      journal = {\apj},
         year = 2019,
        month = nov,
       volume = {885},
       number = {1},
          eid = {57},
        pages = {57},
          doi = {10.3847/1538-4357/ab418f},
archivePrefix = {arXiv},
       eprint = {1909.01368},
 primaryClass = {astro-ph.GA},
       adsurl = {https://ui.adsabs.harvard.edu/abs/2019ApJ...885...57W}
}

@ARTICLE{Izotov21,
       author = {{Izotov}, Y.~I. and {Worseck}, G. and {Schaerer}, D. and {Guseva}, N.~G. and {Chisholm}, J. and {Thuan}, T.~X. and {Fricke}, K.~J. and {Verhamme}, A.},
        title = "{Lyman continuum leakage from low-mass galaxies with M$_{{\ensuremath{\star}}}$ < {}10$^{8}$ M$_{{\ensuremath{\odot}}}$}",
      journal = {\mnras},
         year = 2021,
        month = may,
       volume = {503},
       number = {2},
        pages = {1734-1752},
          doi = {10.1093/mnras/stab612},
archivePrefix = {arXiv},
       eprint = {2103.01514},
 primaryClass = {astro-ph.GA},
       adsurl = {https://ui.adsabs.harvard.edu/abs/2021MNRAS.503.1734I}
}

@BOOK{Osterbrock,
       author = {{Osterbrock}, Donald E. and {Ferland}, Gary J.},
        title = "{Astrophysics of gaseous nebulae and active galactic nuclei}",
         year = 2006,
       adsurl = {https://ui.adsabs.harvard.edu/abs/2006agna.book.....O}
}

@ARTICLE{Garilli21,
       author = {{Garilli}, B. and {McLure}, R. and {Pentericci}, L. and {Franzetti}, P. and {Gargiulo}, A. and {Carnall}, A. and {Cucciati}, O. and {Iovino}, A. and {Amorin}, R. and {Bolzonella}, M. and {Bongiorno}, A. and {Castellano}, M. and {Cimatti}, A. and {Cirasuolo}, M. and {Cullen}, F. and {Dunlop}, J. and {Elbaz}, D. and {Finkelstein}, S. and {Fontana}, A. and {Fontanot}, F. and {Fumana}, M. and {Guaita}, L. and {Hartley}, W. and {Jarvis}, M. and {Juneau}, S. and {Maccagni}, D. and {McLeod}, D. and {Nandra}, K. and {Pompei}, E. and {Pozzetti}, L. and {Scodeggio}, M. and {Talia}, M. and {Calabr{\`o}}, A. and {Cresci}, G. and {Fynbo}, J.~P.~U. and {Hathi}, N.~P. and {Hibon}, P. and {Koekemoer}, A.~M. and {Magliocchetti}, M. and {Salvato}, M. and {Vietri}, G. and {Zamorani}, G. and {Almaini}, O. and {Balestra}, I. and {Bardelli}, S. and {Begley}, R. and {Brammer}, G. and {Bell}, E.~F. and {Bowler}, R.~A.~A. and {Brusa}, M. and {Buitrago}, F. and {Caputi}, C. and {Cassata}, P. and {Charlot}, S. and {Citro}, A. and {Cristiani}, S. and {Curtis-Lake}, E. and {Dickinson}, M. and {Fazio}, G. and {Ferguson}, H.~C. and {Fiore}, F. and {Franco}, M. and {Georgakakis}, A. and {Giavalisco}, M. and {Grazian}, A. and {Hamadouche}, M. and {Jung}, I. and {Kim}, S. and {Khusanova}, Y. and {Le F{\`e}vre}, O. and {Longhetti}, M. and {Lotz}, J. and {Mannucci}, F. and {Maltby}, D. and {Matsuoka}, K. and {Mendez-Hernandez}, H. and {Mendez-Abreu}, J. and {Mignoli}, M. and {Moresco}, M. and {Nonino}, M. and {Pannella}, M. and {Papovich}, C. and {Popesso}, P. and {Roberts-Borsani}, G. and {Rosario}, D.~J. and {Saldana-Lopez}, A. and {Santini}, P. and {Saxena}, A. and {Schaerer}, D. and {Schreiber}, C. and {Stark}, D. and {Tasca}, L.~A.~M. and {Thomas}, R. and {Vanzella}, E. and {Wild}, V. and {Williams}, C. and {Zucca}, E.},
        title = "{The VANDELS ESO public spectroscopic survey. Final data release of 2087 spectra and spectroscopic measurements}",
      journal = {\aap},
         year = 2021,
        month = mar,
       volume = {647},
          eid = {A150},
        pages = {A150},
          doi = {10.1051/0004-6361/202040059},
archivePrefix = {arXiv},
       eprint = {2101.07645},
 primaryClass = {astro-ph.GA},
       adsurl = {https://ui.adsabs.harvard.edu/abs/2021A&A...647A.150G}
}

@ARTICLE{LzLCS_Flury_2,
       author = {{Flury}, Sophia R. and {Jaskot}, Anne E. and {Ferguson}, Harry C. and {Worseck}, G{\'a}bor and {Makan}, Kirill and {Chisholm}, John and {Saldana-Lopez}, Alberto and {Schaerer}, Daniel and {McCandliss}, Stephan R. and {Xu}, Xinfeng and {Wang}, Bingjie and {Oey}, M.~S. and {Ford}, N.~M. and {Heckman}, Timothy and {Ji}, Zhiyuan and {Giavalisco}, Mauro and {Amor{\'\i}n}, Ricardo and {Atek}, Hakim and {Blaizot}, Jeremy and {Borthakur}, Sanchayeeta and {Carr}, Cody and {Castellano}, Marco and {De Barros}, Stephane and {Dickinson}, Mark and {Finkelstein}, Steven L. and {Fleming}, Brian and {Fontanot}, Fabio and {Garel}, Thibault and {Grazian}, Andrea and {Hayes}, Matthew and {Henry}, Alaina and {Mauerhofer}, Valentin and {Micheva}, Genoveva and {Ostlin}, Goran and {Papovich}, Casey and {Pentericci}, Laura and {Ravindranath}, Swara and {Rosdahl}, Joakim and {Rutkowski}, Michael and {Santini}, Paola and {Scarlata}, Claudia and {Teplitz}, Harry and {Thuan}, Trinh and {Trebitsch}, Maxime and {Vanzella}, Eros and {Verhamme}, Anne},
        title = "{The Low-redshift Lyman Continuum Survey. II. New Insights into LyC Diagnostics}",
      journal = {\apj},
         year = 2022,
        month = may,
       volume = {930},
       number = {2},
          eid = {126},
        pages = {126},
          doi = {10.3847/1538-4357/ac61e4},
archivePrefix = {arXiv},
       eprint = {2203.15649},
 primaryClass = {astro-ph.GA},
       adsurl = {https://ui.adsabs.harvard.edu/abs/2022ApJ...930..126F}
}

@ARTICLE{LzLCS_Flury_1,
       author = {{Flury}, Sophia R. and {Jaskot}, Anne E. and {Ferguson}, Harry C. and {Worseck}, G{\'a}bor and {Makan}, Kirill and {Chisholm}, John and {Saldana-Lopez}, Alberto and {Schaerer}, Daniel and {McCandliss}, Stephan and {Wang}, Bingjie and {Ford}, N.~M. and {Heckman}, Timothy and {Ji}, Zhiyuan and {Giavalisco}, Mauro and {Amorin}, Ricardo and {Atek}, Hakim and {Blaizot}, Jeremy and {Borthakur}, Sanchayeeta and {Carr}, Cody and {Castellano}, Marco and {Cristiani}, Stefano and {De Barros}, Stephane and {Dickinson}, Mark and {Finkelstein}, Steven L. and {Fleming}, Brian and {Fontanot}, Fabio and {Garel}, Thibault and {Grazian}, Andrea and {Hayes}, Matthew and {Henry}, Alaina and {Mauerhofer}, Valentin and {Micheva}, Genoveva and {Oey}, M.~S. and {Ostlin}, Goran and {Papovich}, Casey and {Pentericci}, Laura and {Ravindranath}, Swara and {Rosdahl}, Joakim and {Rutkowski}, Michael and {Santini}, Paola and {Scarlata}, Claudia and {Teplitz}, Harry and {Thuan}, Trinh and {Trebitsch}, Maxime and {Vanzella}, Eros and {Verhamme}, Anne and {Xu}, Xinfeng},
        title = "{The Low-redshift Lyman Continuum Survey. I. New, Diverse Local Lyman Continuum Emitters}",
      journal = {\apjs},
         year = 2022,
        month = may,
       volume = {260},
       number = {1},
          eid = {1},
        pages = {1},
          doi = {10.3847/1538-4365/ac5331},
archivePrefix = {arXiv},
       eprint = {2201.11716},
 primaryClass = {astro-ph.GA},
       adsurl = {https://ui.adsabs.harvard.edu/abs/2022ApJS..260....1F}
}

@software{Lephare,
       author = {{Arnouts}, S. and {Ilbert}, O.},
        title = "{LePHARE: Photometric Analysis for Redshift Estimate}",
 howpublished = {Astrophysics Source Code Library, record ascl:1108.009},
         year = 2011,
        month = aug,
          eid = {ascl:1108.009},
       adsurl = {https://ui.adsabs.harvard.edu/abs/2011ascl.soft08009A}
}

@ARTICLE{Saldana-Lopez23,
       author = {{Saldana-Lopez}, A. and {Schaerer}, D. and {Chisholm}, J. and {Calabr{\`o}}, A. and {Pentericci}, L. and {Cullen}, F. and {Saxena}, A. and {Amor{\'\i}n}, R. and {Carnall}, A.~C. and {Fontanot}, F. and {Fynbo}, J.~P.~U. and {Guaita}, L. and {Hathi}, N.~P. and {Hibon}, P. and {Ji}, Z. and {McLeod}, D.~J. and {Pompei}, E. and {Zamorani}, G.},
        title = "{The VANDELS survey: the ionizing properties of star-forming galaxies at 3 {\ensuremath{\leq}} z {\ensuremath{\leq}} 5 using deep rest-frame ultraviolet spectroscopy}",
      journal = {\mnras},
         year = 2023,
        month = jul,
       volume = {522},
       number = {4},
        pages = {6295-6325},
          doi = {10.1093/mnras/stad1283},
archivePrefix = {arXiv},
       eprint = {2211.01351},
 primaryClass = {astro-ph.GA},
       adsurl = {https://ui.adsabs.harvard.edu/abs/2023MNRAS.522.6295S}
}

@ARTICLE{Jennings25,
       author = {{Jennings}, R. Michael and {Henry}, Alaina and {Mauerhofer}, Valentin and {Heckman}, Timothy and {Scarlata}, Claudia and {Carr}, Cody and {Xu}, Xinfeng and {Huberty}, Mason and {Gazagnes}, Simon and {Jaskot}, Anne E. and {Blaizot}, Jeremy and {Verhamme}, Anne and {Flury}, Sophia R. and {Saldana-Lopez}, Alberto and {Hayes}, Matthew J. and {Trebitsch}, Maxime},
        title = "{A Simulated Galaxy Laboratory: Exploring the Observational Effects on UV Spectral Absorption Line Measurements}",
      journal = {\apj},
         year = 2025,
        month = jan,
       volume = {979},
       number = {1},
          eid = {64},
        pages = {64},
          doi = {10.3847/1538-4357/ad9b13},
archivePrefix = {arXiv},
       eprint = {2412.02794},
 primaryClass = {astro-ph.GA},
       adsurl = {https://ui.adsabs.harvard.edu/abs/2025ApJ...979...64J}
}

@ARTICLE{Gazagnes24,
       author = {{Gazagnes}, Simon and {Cullen}, Fergus and {Mauerhofer}, Valentin and {Begley}, Ryan and {Berg}, Danielle and {Blaizot}, Jeremy and {Chisholm}, John and {Garel}, Thibault and {Leclercq}, Floriane and {McLure}, Ross J. and {Verhamme}, Anne},
        title = "{Comparing the VANDELS Sample to a Zoom-in Radiative Hydrodynamical Simulation: Using the Si II and C II Line Spectra as Tracers of Galaxy Evolution and Lyman Continuum Leakage}",
      journal = {\apj},
         year = 2024,
        month = jul,
       volume = {969},
       number = {1},
          eid = {50},
        pages = {50},
          doi = {10.3847/1538-4357/ad47a4},
archivePrefix = {arXiv},
       eprint = {2405.03759},
 primaryClass = {astro-ph.GA},
       adsurl = {https://ui.adsabs.harvard.edu/abs/2024ApJ...969...50G}
}

@ARTICLE{Saldana-Lopez22,
       author = {{Saldana-Lopez}, Alberto and {Schaerer}, Daniel and {Chisholm}, John and {Flury}, Sophia R. and {Jaskot}, Anne E. and {Worseck}, G{\'a}bor and {Makan}, Kirill and {Gazagnes}, Simon and {Mauerhofer}, Valentin and {Verhamme}, Anne and {Amor{\'\i}n}, Ricardo O. and {Ferguson}, Harry C. and {Giavalisco}, Mauro and {Grazian}, Andrea and {Hayes}, Matthew J. and {Heckman}, Timothy M. and {Henry}, Alaina and {Ji}, Zhiyuan and {Marques-Chaves}, Rui and {McCandliss}, Stephan R. and {Oey}, M. Sally and {{\"O}stlin}, G{\"o}ran and {Pentericci}, Laura and {Thuan}, Trinh X. and {Trebitsch}, Maxime and {Vanzella}, Eros and {Xu}, Xinfeng},
        title = "{The Low-Redshift Lyman Continuum Survey. Unveiling the ISM properties of low-z Lyman-continuum emitters}",
      journal = {\aap},
         year = 2022,
        month = jul,
       volume = {663},
          eid = {A59},
        pages = {A59},
          doi = {10.1051/0004-6361/202141864},
archivePrefix = {arXiv},
       eprint = {2201.11800},
 primaryClass = {astro-ph.GA},
       adsurl = {https://ui.adsabs.harvard.edu/abs/2022A&A...663A..59S}
}

@ARTICLE{Choustikov24,
       author = {{Choustikov}, Nicholas and {Katz}, Harley and {Saxena}, Aayush and {Cameron}, Alex J. and {Devriendt}, Julien and {Slyz}, Adrianne and {Rosdahl}, Joki and {Blaizot}, Jeremy and {Michel-Dansac}, Leo},
        title = "{The Physics of Indirect Estimators of Lyman Continuum Escape and their Application to High-Redshift JWST Galaxies}",
      journal = {\mnras},
         year = 2024,
        month = apr,
       volume = {529},
       number = {4},
        pages = {3751-3767},
          doi = {10.1093/mnras/stae776},
archivePrefix = {arXiv},
       eprint = {2304.08526},
 primaryClass = {astro-ph.GA},
       adsurl = {https://ui.adsabs.harvard.edu/abs/2024MNRAS.529.3751C}
}

@ARTICLE{CIGALE,
       author = {{Boquien}, M. and {Burgarella}, D. and {Roehlly}, Y. and {Buat}, V. and {Ciesla}, L. and {Corre}, D. and {Inoue}, A.~K. and {Salas}, H.},
        title = "{CIGALE: a python Code Investigating GALaxy Emission}",
      journal = {\aap},
         year = 2019,
        month = feb,
       volume = {622},
          eid = {A103},
        pages = {A103},
          doi = {10.1051/0004-6361/201834156},
archivePrefix = {arXiv},
       eprint = {1811.03094},
 primaryClass = {astro-ph.GA},
       adsurl = {https://ui.adsabs.harvard.edu/abs/2019A&A...622A.103B}
}

@ARTICLE{Gazagnes23,
       author = {{Gazagnes}, Simon and {Mauerhofer}, Valentin and {Berg}, Danielle A. and {Blaizot}, Jeremy and {Verhamme}, Anne and {Garel}, Thibault and {Erb}, Dawn K. and {Arellano-C{\'o}rdova}, Karla Z. and {Brinchmann}, Jarle and {Chisholm}, John and {Hayes}, Matthew and {Henry}, Alaina and {James}, Bethan L. and {Jaskot}, Anne and {Jurlin}, Nika and {Martin}, Crystal L. and {Maseda}, Michael and {Scarlata}, Claudia and {Skillman}, Evan D. and {Wilkins}, Stephen M. and {Wofford}, Aida and {Xu}, Xinfeng},
        title = "{Interpreting the Si II and C II Line Spectra from the COS Legacy Archive Spectroscopic SurveY Using a Virtual Galaxy from a High-resolution Radiation-hydrodynamic Simulation}",
      journal = {\apj},
         year = 2023,
        month = aug,
       volume = {952},
       number = {2},
          eid = {164},
        pages = {164},
          doi = {10.3847/1538-4357/acda2c},
archivePrefix = {arXiv},
       eprint = {2305.19177},
 primaryClass = {astro-ph.GA},
       adsurl = {https://ui.adsabs.harvard.edu/abs/2023ApJ...952..164G}
}

@ARTICLE{Sphinx20_release,
       author = {{Katz}, Harley and {Rosdahl}, Joki and {Kimm}, Taysun and {Blaizot}, Jeremy and {Choustikov}, Nicholas and {Farcy}, Marion and {Garel}, Thibault and {Haehnelt}, Martin G. and {Michel-Dansac}, Leo and {Ocvirk}, Pierre},
        title = "{The SPHINX Public Data Release: Forward Modelling High-Redshift JWST Observations with Cosmological Radiation Hydrodynamics Simulations}",
      journal = {The Open Journal of Astrophysics},
         year = 2023,
        month = dec,
       volume = {6},
          eid = {44},
        pages = {44},
          doi = {10.21105/astro.2309.03269},
archivePrefix = {arXiv},
       eprint = {2309.03269},
 primaryClass = {astro-ph.GA},
       adsurl = {https://ui.adsabs.harvard.edu/abs/2023OJAp....6E..44K}
}

@ARTICLE{Berg22,
       author = {{Berg}, Danielle A. and {James}, Bethan L. and {King}, Teagan and {McDonald}, Meaghan and {Chen}, Zuyi and {Chisholm}, John and {Heckman}, Timothy and {Martin}, Crystal L. and {Stark}, Dan P. and {Aloisi}, Alessandra and {Amor{\'\i}n}, Ricardo O. and {Arellano-C{\'o}rdova}, Karla Z. and {Bayliss}, Matthew and {Bordoloi}, Rongmon and {Brinchmann}, Jarle and {Charlot}, St{\'e}phane and {Chevallard}, Jacopo and {Clark}, Ilyse and {Erb}, Dawn K. and {Feltre}, Anna and {Gronke}, Max and {Hayes}, Matthew and {Henry}, Alaina and {Hernandez}, Svea and {Jaskot}, Anne and {Jones}, Tucker and {Kewley}, Lisa J. and {Kumari}, Nimisha and {Leitherer}, Claus and {Llerena}, Mario and {Maseda}, Michael and {Mingozzi}, Matilde and {Nanayakkara}, Themiya and {Ouchi}, Masami and {Plat}, Adele and {Pogge}, Richard W. and {Ravindranath}, Swara and {Rigby}, Jane R. and {Sanders}, Ryan and {Scarlata}, Claudia and {Senchyna}, Peter and {Skillman}, Evan D. and {Steidel}, Charles C. and {Strom}, Allison L. and {Sugahara}, Yuma and {Wilkins}, Stephen M. and {Wofford}, Aida and {Xu}, Xinfeng and {Classy Team}},
        title = "{The COS Legacy Archive Spectroscopy Survey (CLASSY) Treasury Atlas}",
      journal = {\apjs},
         year = 2022,
        month = aug,
       volume = {261},
       number = {2},
          eid = {31},
        pages = {31},
          doi = {10.3847/1538-4365/ac6c03},
archivePrefix = {arXiv},
       eprint = {2203.07357},
 primaryClass = {astro-ph.GA},
       adsurl = {https://ui.adsabs.harvard.edu/abs/2022ApJS..261...31B}
}

@ARTICLE{Chisholm22,
       author = {{Chisholm}, J. and {Saldana-Lopez}, A. and {Flury}, S. and {Schaerer}, D. and {Jaskot}, A. and {Amor{\'\i}n}, R. and {Atek}, H. and {Finkelstein}, S.~L. and {Fleming}, B. and {Ferguson}, H. and {Fern{\'a}ndez}, V. and {Giavalisco}, M. and {Hayes}, M. and {Heckman}, T. and {Henry}, A. and {Ji}, Z. and {Marques-Chaves}, R. and {Mauerhofer}, V. and {McCandliss}, S. and {Oey}, M.~S. and {{\"O}stlin}, G. and {Rutkowski}, M. and {Scarlata}, C. and {Thuan}, T. and {Trebitsch}, M. and {Wang}, B. and {Worseck}, G. and {Xu}, X.},
        title = "{The Far-Ultraviolet Continuum Slope as a Lyman Continuum Escape Estimator at High-redshift}",
      journal = {\mnras},
         year = 2022,
        month = oct,
          doi = {10.1093/mnras/stac2874},
archivePrefix = {arXiv},
       eprint = {2207.05771},
 primaryClass = {astro-ph.GA},
       adsurl = {https://ui.adsabs.harvard.edu/abs/2022MNRAS.tmp.2655C}
}

@ARTICLE{Konstantopoulou22,
       author = {{Konstantopoulou}, Christina and {De Cia}, Annalisa and {Krogager}, Jens-Kristian and {Ledoux}, C{\'e}dric and {Noterdaeme}, Pasquier and {Fynbo}, Johan P.~U. and {Heintz}, Kasper E. and {Watson}, Darach and {Andersen}, Anja C. and {Ramburuth-Hurt}, Tanita and {Jermann}, Iris},
        title = "{Dust depletion of metals from local to distant galaxies. I. Peculiar nucleosynthesis effects and grain growth in the ISM}",
      journal = {\aap},
         year = 2022,
        month = oct,
       volume = {666},
          eid = {A12},
        pages = {A12},
          doi = {10.1051/0004-6361/202243994},
archivePrefix = {arXiv},
       eprint = {2207.08804},
 primaryClass = {astro-ph.GA},
       adsurl = {https://ui.adsabs.harvard.edu/abs/2022A&A...666A..12K}
}

@ARTICLE{DeCia16,
       author = {{De Cia}, A. and {Ledoux}, C. and {Mattsson}, L. and {Petitjean}, P. and {Srianand}, R. and {Gavignaud}, I. and {Jenkins}, E.~B.},
        title = "{Dust-depletion sequences in damped Lyman-{\ensuremath{\alpha}} absorbers. A unified picture from low-metallicity systems to the Galaxy}",
      journal = {\aap},
         year = 2016,
        month = dec,
       volume = {596},
          eid = {A97},
        pages = {A97},
          doi = {10.1051/0004-6361/201527895},
archivePrefix = {arXiv},
       eprint = {1608.08621},
 primaryClass = {astro-ph.GA},
       adsurl = {https://ui.adsabs.harvard.edu/abs/2016A&A...596A..97D}
}

@ARTICLE{Katz22,
       author = {{Katz}, Harley and {Rosdahl}, Joakim and {Kimm}, Taysun and {Garel}, Thibault and {Blaizot}, J{\'e}r{\'e}my and {Haehnelt}, Martin G. and {Michel-Dansac}, L{\'e}o and {Martin-Alvarez}, Sergio and {Devriendt}, Julien and {Slyz}, Adrianne and {Teyssier}, Romain and {Ocvirk}, Pierre and {Laporte}, Nicolas and {Ellis}, Richard},
        title = "{The nature of high [O III]$_{88 {\ensuremath{\mu}} m}$/[C II]$_{158 {\ensuremath{\mu}}m}$ galaxies in the epoch of reionization: Low carbon abundance and a top-heavy IMF?}",
      journal = {\mnras},
         year = 2022,
        month = mar,
       volume = {510},
       number = {4},
        pages = {5603-5622},
          doi = {10.1093/mnras/stac028},
archivePrefix = {arXiv},
       eprint = {2108.01074},
 primaryClass = {astro-ph.GA},
       adsurl = {https://ui.adsabs.harvard.edu/abs/2022MNRAS.510.5603K}
}

@ARTICLE{Sphinx20,
       author = {{Rosdahl}, Joakim and {Blaizot}, J{\'e}r{\'e}my and {Katz}, Harley and {Kimm}, Taysun and {Garel}, Thibault and {Haehnelt}, Martin and {Keating}, Laura C. and {Martin-Alvarez}, Sergio and {Michel-Dansac}, L{\'e}o and {Ocvirk}, Pierre},
        title = "{LyC escape from SPHINX galaxies in the Epoch of Reionization}",
      journal = {\mnras},
         year = 2022,
        month = sep,
       volume = {515},
       number = {2},
        pages = {2386-2414},
          doi = {10.1093/mnras/stac1942},
archivePrefix = {arXiv},
       eprint = {2207.03232},
 primaryClass = {astro-ph.GA},
       adsurl = {https://ui.adsabs.harvard.edu/abs/2022MNRAS.515.2386R}
}

@ARTICLE{Mauerhofer21,
       author = {{Mauerhofer}, V. and {Verhamme}, A. and {Blaizot}, J. and {Garel}, T. and {Kimm}, T. and {Michel-Dansac}, L. and {Rosdahl}, J.},
        title = "{UV absorption lines and their potential for tracing the Lyman continuum escape fraction}",
      journal = {\aap},
         year = 2021,
        month = feb,
       volume = {646},
          eid = {A80},
        pages = {A80},
          doi = {10.1051/0004-6361/202039449},
archivePrefix = {arXiv},
       eprint = {2012.03984},
 primaryClass = {astro-ph.GA},
       adsurl = {https://ui.adsabs.harvard.edu/abs/2021A&A...646A..80M}
}

@ARTICLE{Fan06,
   author = {{Fan}, X. and {Carilli}, C.~L. and {Keating}, B.},
    title = "{Observational Constraints on Cosmic Reionization}",
  journal = {\araa},
   eprint = {astro-ph/0602375},
     year = 2006,
    month = sep,
   volume = 44,
    pages = {415-462},
      doi = {10.1146/annurev.astro.44.051905.092514},
   adsurl = {http://adsabs.harvard.edu/abs/2006ARA%26A..44..415F}
}

@ARTICLE{2018PASJ...70S..13O,
   author = {{Ouchi}, M. and {Harikane}, Y. and {Shibuya}, T. and {Shimasaku}, K. and 
	{Taniguchi}, Y. and {Konno}, A. and {Kobayashi}, M. and {Kajisawa}, M. and 
	{Nagao}, T. and {Ono}, Y. and {Inoue}, A.~K. and {Umemura}, M. and 
	{Mori}, M. and {Hasegawa}, K. and {Higuchi}, R. and {Komiyama}, Y. and 
	{Matsuda}, Y. and {Nakajima}, K. and {Saito}, T. and {Wang}, S.-Y.
	},
    title = "{Systematic Identification of LAEs for Visible Exploration and Reionization Research Using Subaru HSC (SILVERRUSH). I. Program strategy and clustering properties of {\tilde}2000 Ly{$\alpha$} emitters at z = 6-7 over the 0.3-0.5 Gpc$^{2}$ survey area}",
  journal = {\pasj},
archivePrefix = "arXiv",
   eprint = {1704.07455},
     year = 2018,
    month = jan,
   volume = 70,
      eid = {S13},
    pages = {S13},
      doi = {10.1093/pasj/psx074},
   adsurl = {http://adsabs.harvard.edu/abs/2018PASJ...70S..13O}
}

@ARTICLE{2014MNRAS.439.2386G,
   author = {{Grassi}, T. and {Bovino}, S. and {Schleicher}, D.~R.~G. and 
	{Prieto}, J. and {Seifried}, D. and {Simoncini}, E. and {Gianturco}, F.~A.
	},
    title = "{KROME - a package to embed chemistry in astrophysical simulations}",
  journal = {\mnras},
archivePrefix = "arXiv",
   eprint = {1311.1070},
     year = 2014,
    month = apr,
   volume = 439,
    pages = {2386-2419},
      doi = {10.1093/mnras/stu114},
   adsurl = {http://adsabs.harvard.edu/abs/2014MNRAS.439.2386G}
}

@ARTICLE{Verner96,
   author = {{Verner}, D.~A. and {Ferland}, G.~J. and {Korista}, K.~T. and 
	{Yakovlev}, D.~G.},
    title = "{Atomic Data for Astrophysics. II. New Analytic FITS for Photoionization Cross Sections of Atoms and Ions}",
  journal = {\apj},
   eprint = {astro-ph/9601009},
     year = 1996,
    month = jul,
   volume = 465,
    pages = {487},
      doi = {10.1086/177435},
   adsurl = {http://adsabs.harvard.edu/abs/1996ApJ...465..487V}
}

@ARTICLE{Voronov97,
   author = {{Voronov}, G.~S.},
    title = "{A Practical Fit Formula for Ionization Rate Coefficients of Atoms and Ions by Electron Impact: Z = 1-28}",
  journal = {Atomic Data and Nuclear Data Tables},
     year = 1997,
   volume = 65,
    pages = {1},
      doi = {10.1006/adnd.1997.0732},
   adsurl = {http://adsabs.harvard.edu/abs/1997ADNDT..65....1V}
}

@ARTICLE{Kimm15,
       author = {{Kimm}, Taysun and {Cen}, Renyue and {Devriendt}, Julien and {Dubois}, Yohan and {Slyz}, Adrianne},
        title = "{Towards simulating star formation in turbulent high-z galaxies with mechanical supernova feedback}",
      journal = {\mnras},
         year = 2015,
        month = aug,
       volume = {451},
       number = {3},
        pages = {2900-2921},
          doi = {10.1093/mnras/stv1211},
archivePrefix = {arXiv},
       eprint = {1501.05655},
 primaryClass = {astro-ph.GA},
       adsurl = {https://ui.adsabs.harvard.edu/abs/2015MNRAS.451.2900K}
}

@ARTICLE{BadnellRR,
   author = {{Badnell}, N.~R.},
    title = "{Radiative Recombination Data for Modeling Dynamic Finite-Density Plasmas}",
  journal = {\apjs},
   eprint = {astro-ph/0604144},
     year = 2006,
    month = dec,
   volume = 167,
    pages = {334-342},
      doi = {10.1086/508465},
   adsurl = {http://adsabs.harvard.edu/abs/2006ApJS..167..334B}
}

@ARTICLE{Joki2013,
       author = {{Rosdahl}, J. and {Blaizot}, J. and {Aubert}, D. and {Stranex}, T. and
         {Teyssier}, R.},
        title = "{RAMSES-RT: radiation hydrodynamics in the cosmological context}",
      journal = {\mnras},
         year = "2013",
        month = "Dec",
       volume = {436},
       number = {3},
        pages = {2188-2231},
          doi = {10.1093/mnras/stt1722},
archivePrefix = {arXiv},
       eprint = {1304.7126},
 primaryClass = {astro-ph.CO},
       adsurl = {https://ui.adsabs.harvard.edu/abs/2013MNRAS.436.2188R}
}

@ARTICLE{Sphinx,
       author = {{Rosdahl}, Joakim and {Katz}, Harley and {Blaizot}, J{\'e}r{\'e}my and
         {Kimm}, Taysun and {Michel-Dansac}, L{\'e}o and {Garel}, Thibault and
         {Haehnelt}, Martin and {Ocvirk}, Pierre and {Teyssier}, Romain},
        title = "{The SPHINX cosmological simulations of the first billion years: the impact of binary stars on reionization}",
      journal = {\mnras},
         year = "2018",
        month = "Sep",
       volume = {479},
       number = {1},
        pages = {994-1016},
          doi = {10.1093/mnras/sty1655},
archivePrefix = {arXiv},
       eprint = {1801.07259},
 primaryClass = {astro-ph.GA},
       adsurl = {https://ui.adsabs.harvard.edu/abs/2018MNRAS.479..994R}
}

@ARTICLE{BPASS1,
       author = {{Eldridge}, John J. and {Izzard}, Robert G. and {Tout}, Christopher A.},
        title = "{The effect of massive binaries on stellar populations and supernova progenitors}",
      journal = {\mnras},
         year = "2008",
        month = "Mar",
       volume = {384},
       number = {3},
        pages = {1109-1118},
          doi = {10.1111/j.1365-2966.2007.12738.x},
archivePrefix = {arXiv},
       eprint = {0711.3079},
 primaryClass = {astro-ph},
       adsurl = {https://ui.adsabs.harvard.edu/abs/2008MNRAS.384.1109E}
}

@ARTICLE{BPASS2,
       author = {{Stanway}, Elizabeth R. and {Eldridge}, J.~J. and {Becker}, George D.},
        title = "{Stellar population effects on the inferred photon density at reionization}",
      journal = {\mnras},
         year = "2016",
        month = "Feb",
       volume = {456},
       number = {1},
        pages = {485-499},
          doi = {10.1093/mnras/stv2661},
archivePrefix = {arXiv},
       eprint = {1511.03268},
 primaryClass = {astro-ph.GA},
       adsurl = {https://ui.adsabs.harvard.edu/abs/2016MNRAS.456..485S}
}

@ARTICLE{Laursen09b,
       author = {{Laursen}, Peter and {Sommer-Larsen}, Jesper and {Andersen}, Anja C.},
        title = "{Ly{\ensuremath{\alpha}} Radiative Transfer with Dust: Escape Fractions from Simulated High-Redshift Galaxies}",
      journal = {\apj},
         year = "2009",
        month = "Oct",
       volume = {704},
       number = {2},
        pages = {1640-1656},
          doi = {10.1088/0004-637X/704/2/1640},
archivePrefix = {arXiv},
       eprint = {0907.2698},
 primaryClass = {astro-ph.CO},
       adsurl = {https://ui.adsabs.harvard.edu/abs/2009ApJ...704.1640L}
}

@ARTICLE{Li01,
       author = {{Li}, Aigen and {Draine}, B.~T.},
        title = "{Infrared Emission from Interstellar Dust. II. The Diffuse Interstellar Medium}",
      journal = {\apj},
         year = "2001",
        month = "Jun",
       volume = {554},
       number = {2},
        pages = {778-802},
          doi = {10.1086/323147},
archivePrefix = {arXiv},
       eprint = {astro-ph/0011319},
 primaryClass = {astro-ph},
       adsurl = {https://ui.adsabs.harvard.edu/abs/2001ApJ...554..778L}
}

@ARTICLE{Gnedin08,
       author = {{Gnedin}, Nickolay Y. and {Kravtsov}, Andrey V. and {Chen}, Hsiao-Wen},
        title = "{Escape of Ionizing Radiation from High-Redshift Galaxies}",
      journal = {\apj},
         year = "2008",
        month = "Jan",
       volume = {672},
       number = {2},
        pages = {765-775},
          doi = {10.1086/524007},
archivePrefix = {arXiv},
       eprint = {0707.0879},
 primaryClass = {astro-ph},
       adsurl = {https://ui.adsabs.harvard.edu/abs/2008ApJ...672..765G}
}

@ARTICLE{Gazagnes18,
       author = {{Gazagnes}, S. and {Chisholm}, J. and {Schaerer}, D. and {Verhamme}, A. and
         {Rigby}, J.~R. and {Bayliss}, M.},
        title = "{Neutral gas properties of Lyman continuum emitting galaxies: Column densities and covering fractions from UV absorption lines}",
      journal = {\aap},
         year = "2018",
        month = "Aug",
       volume = {616},
          eid = {A29},
        pages = {A29},
          doi = {10.1051/0004-6361/201832759},
archivePrefix = {arXiv},
       eprint = {1802.06378},
 primaryClass = {astro-ph.GA},
       adsurl = {https://ui.adsabs.harvard.edu/abs/2018A&A...616A..29G}
}

@ARTICLE{Smith_Colt,
       author = {{Smith}, Aaron and {Safranek-Shrader}, Chalence and {Bromm}, Volker and
         {Milosavljevi{\'c}}, Milo{\v{s}}},
        title = "{The Lyman {\ensuremath{\alpha}} signature of the first galaxies}",
      journal = {\mnras},
         year = "2015",
        month = "Jun",
       volume = {449},
       number = {4},
        pages = {4336-4362},
          doi = {10.1093/mnras/stv565},
archivePrefix = {arXiv},
       eprint = {1409.4480},
 primaryClass = {astro-ph.CO},
       adsurl = {https://ui.adsabs.harvard.edu/abs/2015MNRAS.449.4336S}
}

@ARTICLE{HM12,
       author = {{Haardt}, Francesco and {Madau}, Piero},
        title = "{Radiative Transfer in a Clumpy Universe. IV. New Synthesis Models of the Cosmic UV/X-Ray Background}",
      journal = {\apj},
         year = "2012",
        month = "Feb",
       volume = {746},
       number = {2},
          eid = {125},
        pages = {125},
          doi = {10.1088/0004-637X/746/2/125},
archivePrefix = {arXiv},
       eprint = {1105.2039},
 primaryClass = {astro-ph.CO},
       adsurl = {https://ui.adsabs.harvard.edu/abs/2012ApJ...746..125H}
}

@ARTICLE{Heckman11,
       author = {{Heckman}, Timothy M. and {Borthakur}, Sanchayeeta and
         {Overzier}, Roderik and {Kauffmann}, Guinevere and {Basu-Zych}, Antara and
         {Leitherer}, Claus and {Sembach}, Ken and {Martin}, D. Chris and
         {Rich}, R. Michael and {Schiminovich}, David and {Seibert}, Mark},
        title = "{Extreme Feedback and the Epoch of Reionization: Clues in the Local Universe}",
      journal = {\apj},
         year = "2011",
        month = "Mar",
       volume = {730},
       number = {1},
          eid = {5},
        pages = {5},
          doi = {10.1088/0004-637X/730/1/5},
archivePrefix = {arXiv},
       eprint = {1101.4219},
 primaryClass = {astro-ph.CO},
       adsurl = {https://ui.adsabs.harvard.edu/abs/2011ApJ...730....5H}
}

@ARTICLE{Chisholm17,
       author = {{Chisholm}, J. and {Orlitov{\'a}}, I. and {Schaerer}, D. and
         {Verhamme}, A. and {Worseck}, G. and {Izotov}, Y.~I. and
         {Thuan}, T.~X. and {Guseva}, N.~G.},
        title = "{Do galaxies that leak ionizing photons have extreme outflows?}",
      journal = {\aap},
         year = "2017",
        month = "Sep",
       volume = {605},
          eid = {A67},
        pages = {A67},
          doi = {10.1051/0004-6361/201730610},
archivePrefix = {arXiv},
       eprint = {1707.01913},
 primaryClass = {astro-ph.GA},
       adsurl = {https://ui.adsabs.harvard.edu/abs/2017A&A...605A..67C}
}

@ARTICLE{Chisholm18,
       author = {{Chisholm}, J. and {Gazagnes}, S. and {Schaerer}, D. and {Verhamme}, A. and
         {Rigby}, J.~R. and {Bayliss}, M. and {Sharon}, K. and {Gladders}, M. and
         {Dahle}, H.},
        title = "{Accurately predicting the escape fraction of ionizing photons using rest-frame ultraviolet absorption lines}",
      journal = {\aap},
         year = "2018",
        month = "Aug",
       volume = {616},
          eid = {A30},
        pages = {A30},
          doi = {10.1051/0004-6361/201832758},
archivePrefix = {arXiv},
       eprint = {1803.03655},
 primaryClass = {astro-ph.GA},
       adsurl = {https://ui.adsabs.harvard.edu/abs/2018A&A...616A..30C}
}

@ARTICLE{Borthakur14,
       author = {{Borthakur}, Sanchayeeta and {Heckman}, Timothy M. and
         {Leitherer}, Claus and {Overzier}, Roderik A.},
        title = "{A local clue to the reionization of the universe}",
      journal = {Science},
         year = "2014",
        month = "Oct",
       volume = {346},
       number = {6206},
        pages = {216-219},
          doi = {10.1126/science.1254214},
archivePrefix = {arXiv},
       eprint = {1410.3511},
 primaryClass = {astro-ph.GA},
       adsurl = {https://ui.adsabs.harvard.edu/abs/2014Sci...346..216B}
}

@ARTICLE{Pyneb,
       author = {{Luridiana}, V. and {Morisset}, C. and {Shaw}, R.~A.},
        title = "{PyNeb: a new tool for analyzing emission lines. I. Code description and validation of results}",
      journal = {\aap},
         year = "2015",
        month = "Jan",
       volume = {573},
          eid = {A42},
        pages = {A42},
          doi = {10.1051/0004-6361/201323152},
archivePrefix = {arXiv},
       eprint = {1410.6662},
 primaryClass = {astro-ph.IM},
       adsurl = {https://ui.adsabs.harvard.edu/abs/2015A&A...573A..42L}
}

@ARTICLE{Henyey41,
       author = {{Henyey}, L.~G. and {Greenstein}, J.~L.},
        title = "{Diffuse radiation in the Galaxy.}",
      journal = {\apj},
         year = "1941",
        month = "Jan",
       volume = {93},
        pages = {70-83},
          doi = {10.1086/144246},
       adsurl = {https://ui.adsabs.harvard.edu/abs/1941ApJ....93...70H}
}

@ARTICLE{Reddy16,
       author = {{Reddy}, Naveen A. and {Steidel}, Charles C. and {Pettini}, Max and
         {Bogosavljevi{\'c}}, Milan and {Shapley}, Alice E.},
        title = "{The Connection Between Reddening, Gas Covering Fraction, and the Escape of Ionizing Radiation at High Redshift}",
      journal = {\apj},
         year = "2016",
        month = "Sep",
       volume = {828},
       number = {2},
          eid = {108},
        pages = {108},
          doi = {10.3847/0004-637X/828/2/108},
archivePrefix = {arXiv},
       eprint = {1606.03452},
 primaryClass = {astro-ph.GA},
       adsurl = {https://ui.adsabs.harvard.edu/abs/2016ApJ...828..108R}
}

@ARTICLE{Shapley03,
       author = {{Shapley}, Alice E. and {Steidel}, Charles C. and {Pettini}, Max and
         {Adelberger}, Kurt L.},
        title = "{Rest-Frame Ultraviolet Spectra of z\raisebox{-0.5ex}\textasciitilde3 Lyman Break Galaxies}",
      journal = {\apj},
         year = "2003",
        month = "May",
       volume = {588},
       number = {1},
        pages = {65-89},
          doi = {10.1086/373922},
archivePrefix = {arXiv},
       eprint = {astro-ph/0301230},
 primaryClass = {astro-ph},
       adsurl = {https://ui.adsabs.harvard.edu/abs/2003ApJ...588...65S}
}

@ARTICLE{Steidel18,
       author = {{Steidel}, Charles C. and {Bogosavljevi{\'c}}, Milan and
         {Shapley}, Alice E. and {Reddy}, Naveen A. and {Rudie}, Gwen C. and
         {Pettini}, Max and {Trainor}, Ryan F. and {Strom}, Allison L.},
        title = "{The Keck Lyman Continuum Spectroscopic Survey (KLCS): The Emergent Ionizing Spectrum of Galaxies at z ̃ 3}",
      journal = {\apj},
         year = "2018",
        month = "Dec",
       volume = {869},
       number = {2},
          eid = {123},
        pages = {123},
          doi = {10.3847/1538-4357/aaed28},
archivePrefix = {arXiv},
       eprint = {1805.06071},
 primaryClass = {astro-ph.GA},
       adsurl = {https://ui.adsabs.harvard.edu/abs/2018ApJ...869..123S}
}

@ARTICLE{Steidel10,
       author = {{Steidel}, Charles C. and {Erb}, Dawn K. and {Shapley}, Alice E. and
         {Pettini}, Max and {Reddy}, Naveen and {Bogosavljevi{\'c}}, Milan and
         {Rudie}, Gwen C. and {Rakic}, Olivera},
        title = "{The Structure and Kinematics of the Circumgalactic Medium from Far-ultraviolet Spectra of z \raisebox{-0.5ex}\textasciitilde= 2-3 Galaxies}",
      journal = {\apj},
         year = "2010",
        month = "Jul",
       volume = {717},
       number = {1},
        pages = {289-322},
          doi = {10.1088/0004-637X/717/1/289},
archivePrefix = {arXiv},
       eprint = {1003.0679},
 primaryClass = {astro-ph.CO},
       adsurl = {https://ui.adsabs.harvard.edu/abs/2010ApJ...717..289S}
}

@ARTICLE{Leitherer99,
       author = {{Leitherer}, Claus and {Schaerer}, Daniel and {Goldader}, Jeffrey D. and {Delgado}, Rosa M. Gonz{\'a}lez and {Robert}, Carmelle and {Kune}, Denis Foo and {de Mello}, Du{\'\i}lia F. and {Devost}, Daniel and {Heckman}, Timothy M.},
        title = "{Starburst99: Synthesis Models for Galaxies with Active Star Formation}",
      journal = {\apjs},
         year = 1999,
        month = jul,
       volume = {123},
       number = {1},
        pages = {3-40},
          doi = {10.1086/313233},
archivePrefix = {arXiv},
       eprint = {astro-ph/9902334},
 primaryClass = {astro-ph},
       adsurl = {https://ui.adsabs.harvard.edu/abs/1999ApJS..123....3L}
}

@ARTICLE{Izotov16b,
       author = {{Izotov}, Y.~I. and {Schaerer}, D. and {Thuan}, T.~X. and {Worseck}, G. and
         {Guseva}, N.~G. and {Orlitov{\'a}}, I. and {Verhamme}, A.},
        title = "{Detection of high Lyman continuum leakage from four low-redshift compact star-forming galaxies}",
      journal = {\mnras},
         year = "2016",
        month = "Oct",
       volume = {461},
       number = {4},
        pages = {3683-3701},
          doi = {10.1093/mnras/stw1205},
archivePrefix = {arXiv},
       eprint = {1605.05160},
 primaryClass = {astro-ph.GA},
       adsurl = {https://ui.adsabs.harvard.edu/abs/2016MNRAS.461.3683I}
}

@ARTICLE{Izotov18b,
       author = {{Izotov}, Y.~I. and {Worseck}, G. and {Schaerer}, D. and
         {Guseva}, N.~G. and {Thuan}, T.~X. and {Fricke}, Verhamme, A. and
         {Orlitov{\'a}}, I.},
        title = "{Low-redshift Lyman continuum leaking galaxies with high [O III]/[O II] ratios}",
      journal = {\mnras},
         year = "2018",
        month = "Aug",
       volume = {478},
       number = {4},
        pages = {4851-4865},
          doi = {10.1093/mnras/sty1378},
archivePrefix = {arXiv},
       eprint = {1805.09865},
 primaryClass = {astro-ph.GA},
       adsurl = {https://ui.adsabs.harvard.edu/abs/2018MNRAS.478.4851I}
}

@ARTICLE{Rascas,
       author = {{Michel-Dansac}, L. and {Blaizot}, J. and {Garel}, T. and
         {Verhamme}, A. and {Kimm}, T. and {Trebitsch}, M.},
        title = "{RASCAS: RAdiation SCattering in Astrophysical Simulations}",
      journal = {\aap},
         year = 2020,
        month = mar,
       volume = {635},
          eid = {A154},
        pages = {A154},
          doi = {10.1051/0004-6361/201834961},
archivePrefix = {arXiv},
       eprint = {2001.11252},
 primaryClass = {astro-ph.GA},
       adsurl = {https://ui.adsabs.harvard.edu/abs/2020A&A...635A.154M}
}

@ARTICLE{Scarlata15,
       author = {{Scarlata}, C. and {Panagia}, N.},
        title = "{A Semi-analytical Line Transfer Model to Interpret the Spectra of Galaxy Outflows}",
      journal = {\apj},
         year = 2015,
        month = mar,
       volume = {801},
       number = {1},
          eid = {43},
        pages = {43},
          doi = {10.1088/0004-637X/801/1/43},
archivePrefix = {arXiv},
       eprint = {1501.07282},
 primaryClass = {astro-ph.GA},
       adsurl = {https://ui.adsabs.harvard.edu/abs/2015ApJ...801...43S}
}

@ARTICLE{Heckman15,
       author = {{Heckman}, Timothy M. and {Alexandroff}, Rachel M. and
         {Borthakur}, Sanchayeeta and {Overzier}, Roderik and {Leitherer}, Claus},
        title = "{The Systematic Properties of the Warm Phase of Starburst-Driven Galactic Winds}",
      journal = {\apj},
         year = 2015,
        month = aug,
       volume = {809},
       number = {2},
          eid = {147},
        pages = {147},
          doi = {10.1088/0004-637X/809/2/147},
archivePrefix = {arXiv},
       eprint = {1507.05622},
 primaryClass = {astro-ph.GA},
       adsurl = {https://ui.adsabs.harvard.edu/abs/2015ApJ...809..147H}
}

@ARTICLE{Ramses,
       author = {{Teyssier}, R.},
        title = "{Cosmological hydrodynamics with adaptive mesh refinement. A new high resolution code called RAMSES}",
      journal = {\aap},
         year = 2002,
        month = apr,
       volume = {385},
        pages = {337-364},
          doi = {10.1051/0004-6361:20011817},
archivePrefix = {arXiv},
       eprint = {astro-ph/0111367},
 primaryClass = {astro-ph},
       adsurl = {https://ui.adsabs.harvard.edu/abs/2002A&A...385..337T}
}

@ARTICLE{Erb15,
       author = {{Erb}, Dawn K.},
        title = "{Feedback in low-mass galaxies in the early Universe}",
      journal = {\nat},
         year = 2015,
        month = jul,
       volume = {523},
       number = {7559},
        pages = {169-176},
          doi = {10.1038/nature14454},
archivePrefix = {arXiv},
       eprint = {1507.02374},
 primaryClass = {astro-ph.GA},
       adsurl = {https://ui.adsabs.harvard.edu/abs/2015Natur.523..169E}
}

@ARTICLE{Prochaska11,
       author = {{Prochaska}, J. Xavier and {Kasen}, Daniel and {Rubin}, Kate},
        title = "{Simple Models of Metal-line Absorption and Emission from Cool Gas Outflows}",
      journal = {\apj},
         year = 2011,
        month = jun,
       volume = {734},
       number = {1},
          eid = {24},
        pages = {24},
          doi = {10.1088/0004-637X/734/1/24},
archivePrefix = {arXiv},
       eprint = {1102.3444},
 primaryClass = {astro-ph.GA},
       adsurl = {https://ui.adsabs.harvard.edu/abs/2011ApJ...734...24P}
}

@ARTICLE{Verhamme17,
       author = {{Verhamme}, A. and {Orlitov{\'a}}, I. and {Schaerer}, D. and
         {Izotov}, Y. and {Worseck}, G. and {Thuan}, T.~X. and {Guseva}, N.},
        title = "{Lyman-{\ensuremath{\alpha}} spectral properties of five newly discovered Lyman continuum emitters}",
      journal = {\aap},
         year = 2017,
        month = jan,
       volume = {597},
          eid = {A13},
        pages = {A13},
          doi = {10.1051/0004-6361/201629264},
archivePrefix = {arXiv},
       eprint = {1609.03477},
 primaryClass = {astro-ph.GA},
       adsurl = {https://ui.adsabs.harvard.edu/abs/2017A&A...597A..13V}
}

@ARTICLE{Izotov16a,
       author = {{Izotov}, Y.~I. and {Orlitov{\'a}}, I. and {Schaerer}, D. and
         {Thuan}, T.~X. and {Verhamme}, A. and {Guseva}, N.~G. and {Worseck}, G.},
        title = "{Eight per cent leakage of Lyman continuum photons from a compact, star-forming dwarf galaxy}",
      journal = {\nat},
         year = 2016,
        month = jan,
       volume = {529},
       number = {7585},
        pages = {178-180},
          doi = {10.1038/nature16456},
archivePrefix = {arXiv},
       eprint = {1601.03068},
 primaryClass = {astro-ph.GA},
       adsurl = {https://ui.adsabs.harvard.edu/abs/2016Natur.529..178I}
}

@ARTICLE{Carr18,
       author = {{Carr}, Cody and {Scarlata}, Claudia and {Panagia}, Nino and
         {Henry}, Alaina},
        title = "{A Semi-analytical Line Transfer (SALT) Model. II: The Effects of a Bi-conical Geometry}",
      journal = {\apj},
         year = 2018,
        month = jun,
       volume = {860},
       number = {2},
          eid = {143},
        pages = {143},
          doi = {10.3847/1538-4357/aac48e},
archivePrefix = {arXiv},
       eprint = {1805.05981},
 primaryClass = {astro-ph.GA},
       adsurl = {https://ui.adsabs.harvard.edu/abs/2018ApJ...860..143C}
}

@ARTICLE{Bergvall06,
       author = {{Bergvall}, N. and {Zackrisson}, E. and {Andersson}, B. -G. and
         {Arnberg}, D. and {Masegosa}, J. and {{\"O}stlin}, G.},
        title = "{First detection of Lyman continuum escape from a local starburst galaxy. I. Observations of the luminous blue compact galaxy Haro 11 with the Far Ultraviolet Spectroscopic Explorer (FUSE)}",
      journal = {\aap},
         year = 2006,
        month = mar,
       volume = {448},
       number = {2},
        pages = {513-524},
          doi = {10.1051/0004-6361:20053788},
       adsurl = {https://ui.adsabs.harvard.edu/abs/2006A&A...448..513B}
}

@ARTICLE{Leitet13,
       author = {{Leitet}, E. and {Bergvall}, N. and {Hayes}, M. and {Linn{\'e}}, S. and
         {Zackrisson}, E.},
        title = "{Escape of Lyman continuum radiation from local galaxies. Detection of leakage from the young starburst Tol 1247-232}",
      journal = {\aap},
         year = 2013,
        month = may,
       volume = {553},
          eid = {A106},
        pages = {A106},
          doi = {10.1051/0004-6361/201118370},
archivePrefix = {arXiv},
       eprint = {1302.6971},
 primaryClass = {astro-ph.CO},
       adsurl = {https://ui.adsabs.harvard.edu/abs/2013A&A...553A.106L}
}

@ARTICLE{Leitherer16,
       author = {{Leitherer}, Claus and {Hernandez}, Svea and {Lee}, Janice C. and
         {Oey}, M.~S.},
        title = "{Direct Detection of Lyman Continuum Escape from Local Starburst Galaxies with the Cosmic Origins Spectrograph}",
      journal = {\apj},
         year = 2016,
        month = may,
       volume = {823},
       number = {1},
          eid = {64},
        pages = {64},
          doi = {10.3847/0004-637X/823/1/64},
archivePrefix = {arXiv},
       eprint = {1603.06779},
 primaryClass = {astro-ph.GA},
       adsurl = {https://ui.adsabs.harvard.edu/abs/2016ApJ...823...64L}
}

@ARTICLE{Puschnig17,
       author = {{Puschnig}, J. and {Hayes}, M. and {{\"O}stlin}, G. and
         {Rivera-Thorsen}, T.~E. and {Melinder}, J. and {Cannon}, J.~M. and
         {Menacho}, V. and {Zackrisson}, E. and {Bergvall}, N. and {Leitet}, E.},
        title = "{The Lyman continuum escape and ISM properties in Tololo 1247-232 - new insights from HST and VLA$^{★}$}",
      journal = {\mnras},
         year = 2017,
        month = aug,
       volume = {469},
       number = {3},
        pages = {3252-3269},
          doi = {10.1093/mnras/stx951},
archivePrefix = {arXiv},
       eprint = {1704.05943},
 primaryClass = {astro-ph.GA},
       adsurl = {https://ui.adsabs.harvard.edu/abs/2017MNRAS.469.3252P}
}
\bibliographystyle{aa} % style aa.bst

%APPENDICES
%/////////////////////////////////////////////////////////////////////////////////////////////////////////////////////////////////////////////////////
\begin{appendix}

\section{$\siii$ 1260 plots} \label{sec:1260_plots}

For completeness, we show here the equivalent of Fig.\ref{fig:SiII_structure} and \ref{fig:rflux_old_paper} but using $\siiiline$ instead of $\siiilineb$. In Fig.\ref{fig:SiII1260_structure} we show that the atomic structure related to $\siiiline$ is slightly more complex than for $\siiilineb$, with the upper level being split into two fine-structure levels. All of the transitions displayed are included in our {\sc rascas} modelling. 

\begin{figure}
  \resizebox{\hsize}{!}{\includegraphics{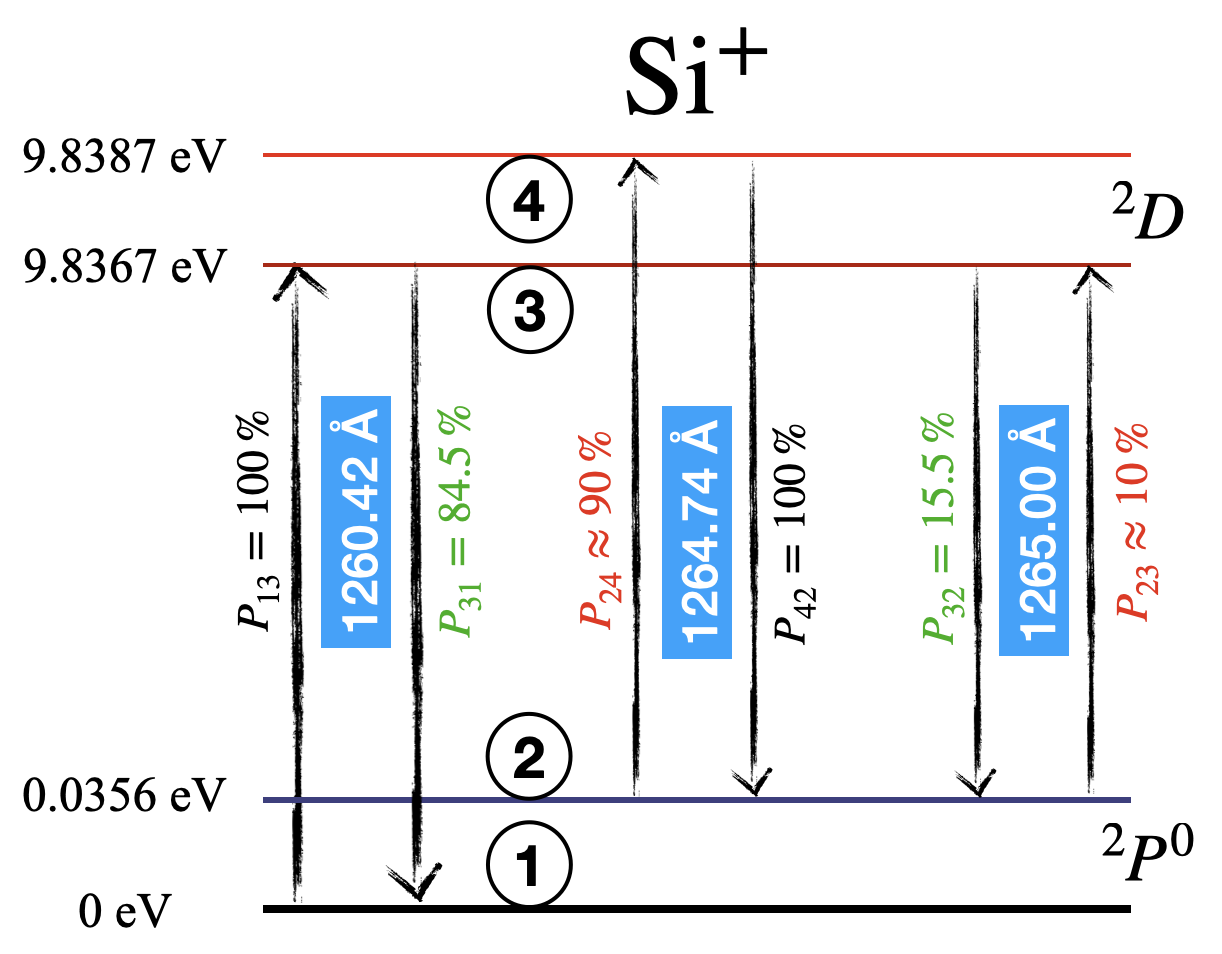}}
  \caption{Energy levels of the $\Siplus$ ion and transitions of $\siiiline$. $P_{13}=100\%$ indicates that a $\Siplus$ ion in level 1 can be photo-excited only to level 3. The green numbers are the probabilities that a $\Siplus$ ion in level 3 radiatively de-excites to level 1 or 2. The red numbers are the probabilities that an incoming photon with a wavelength between $1264.74 \, \angstrom$ and $1265.0 \, \angstrom$ hitting a $\Siplus$ ion in level 2 excites it to level 3 or 4. $P_{42}=100\%$ indicates that a $\Siplus$ ion in level 4 can only radiatively de-excite to level 2.
  }
  \label{fig:SiII1260_structure}
\end{figure}

Then, in Fig.\ref{fig:rflux1260_old_paper}, we show the relation between the dust-corrected residual flux of $\siiiline$ (called $\tilde{R}_{1260}$) and the escape fraction. Contrarily to Fig.\ref{fig:rflux_old_paper}, we use here $\rm{A}_{1300}$ instead of $\rm{A}_{1500}$, since it is closer in wavelength to $\siiiline$. It is clear that using $\siiiline$ results in a wider dispersion. The mean error is 0.0218, which is $42\%$ larger than when using $\siiilineb$ (see Table \ref{tab:fesc_errors}). This explains why we use the latter line in the analysis of this paper.

\begin{figure}
  \resizebox{\hsize}{!}{\includegraphics{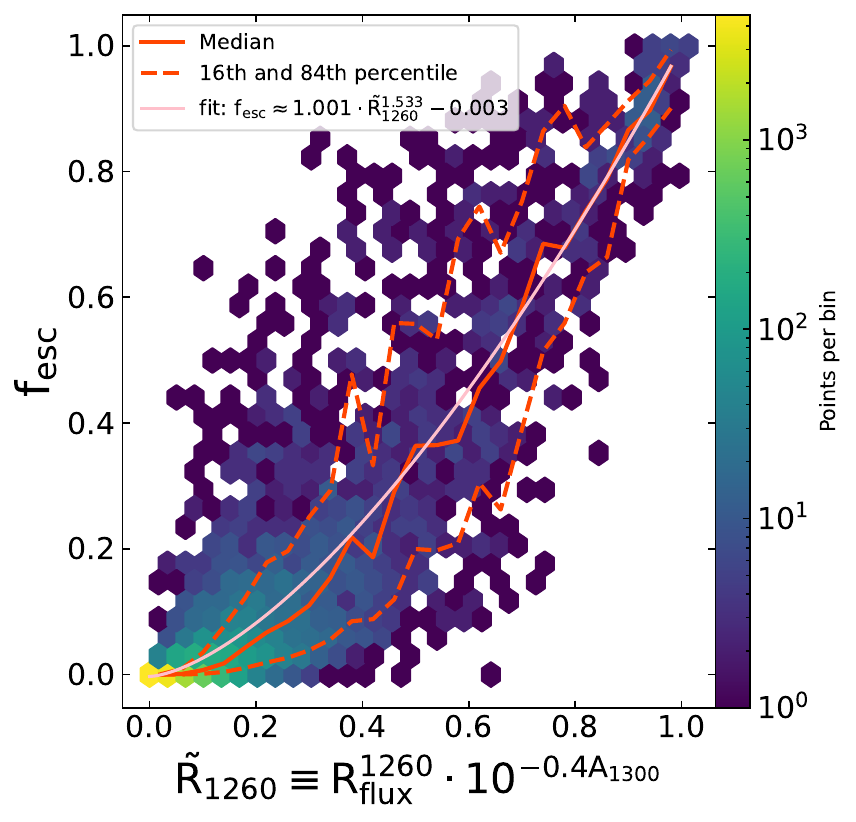}}
  \caption{Comparison of the escape fraction of ionising photons with the product of $\siiiline$ residual flux and the dust attenuation (i.e. the dust-corrected residual flux). The colour scale shows the number of spectra in each hexagonal bin. The solid orange line shows the running median while the dashed lines show the $\sixteenth$ and $\eightyfourth$ percentiles. The pink line shows the best power function fit to the data.
  }
  \label{fig:rflux1260_old_paper}
\end{figure}

\section{Content of the LIS data release} \label{sec:release_content}

The tables and the spectra in JSON format are freely available at the following URL: \url{https://doi.org/10.5281/zenodo.17723577}. A dedicated Jupyter notebook is also provided to illustrate how to read the files and plot example spectra. 

All the data about line properties and dust attenuation is in the table \textit{line\_properties.csv}. This table lists, for each simulated halo (\textit{halo\_id}) and its corresponding redshift, the properties of the $\siii$ absorption and fluorescent emission lines measured along ten different viewing directions (\textit{dir\_0–dir\_9}). For the $\siiiline$ and $\siiilineb$ absorption lines, the table includes the residual flux (\textit{res\_flux}), equivalent width (\textit{EW}), velocity at minimum flux (\textit{v\_max}), and centroid velocity (\textit{v\_cen}). The $\rm{Si \,  \textsc{ii} \, \lambda} 1265$ and $\rm{Si \,  \textsc{ii} \, \lambda} 1533$ entries correspond to their respective fluorescent emission counterparts, with only the equivalent widths recorded. For each LOS, the table also provides the UV attenuation at 1500$\angstrom$ (\textit{$A_{1500}$}), both the value directly measured from the simulation and the values inferred through SED-fitting or spectral-fitting approaches with \textsc{Cigale}, \textsc{lephare}, and \textsc{ficus}. More halo properties are in the original data release \citep{Sphinx20_release}, in the \textit{all\_basic\_data.csv} table.

Then, data about the escape fractions computed in this work is in the table \textit{fesc\_table.csv}. This table provides, for each simulated halo (\textit{halo\_id}) and its corresponding redshift, the escape fraction of ionising photons (\textit{$\fesc$}) measured along ten different viewing directions (\textit{dir\_0–dir\_9}). For each direction, two quantities are reported: the total escape fraction integrated over all ionising wavelengths (\textit{fesc\_dir\_X}) and the escape fraction measured specifically at 900 $\angstrom$ (\textit{fesc\_900\_dir\_X}). The column \textit{fesc\_angle\_avg} gives the angle-averaged escape fraction, defined as the ratio between the total number of ionising photons escaping the galaxy in all directions and the total number intrinsically produced. The quantity \textit{intrinsic\_nion} corresponds to the intrinsic ionising photon production rate of the galaxy (in photons $\rm{s}^{-1}$). Similarly, \textit{intrinsic\_nion\_900} is the intrinsic production rate of photons at $900 \angstrom$, in photons $\rm{s}^{-1} \, \angstrom^{-1}$.

Finally, the data release also includes the synthetic $\siii$ spectra for all galaxies at all seven redshifts. Each file (e.g. SiII\_1260\_spectra\_z6.json) contains the spectra along ten viewing directions for all galaxies at the corresponding redshift. The $\siiiline$ and $\siiilineb$ files include both the absorption and associated fluorescent emission lines within the same spectrum.

\section{Predicting A$_{1500}$ with LEPHARE} \label{sec:lephare}

For comparison with Section \ref{sec:cigale}, where we used \textsc{Cigale}, here we test another independent SED-fitting tool based on photometric measurements, \textsc{lephare} \citep{Lephare}\footnote{\url{https://gitlab.lam.fr/Galaxies/LEPHARE}}. We build SFH models using our stellar library {\sc BPASSv2.2.1} and assume a delayed exponentially declining star-formation history. For the dust attenuation, we choose the SMC extinction law to match our {\sc rascas} models. Like for \textsc{Cigale}, we removed the bluest filter for galaxies at redshifts 4.64 and 5 (see Sect. \ref{sec:cigale}). The resulting predictions of the attenuation $\rm{A_{1500}}$ are shown in Fig.\ref{fig:A1500_lephare}.
Globally, with our setups, \textsc{lephare} predictions are similarly accurate as with \textsc{Cigale}, as shown in Table \ref{tab:fesc_errors}. Attenuations above $\sim 1.5$ mag are on average underestimated. The average error on $\fesc$ is 0.0215 (0.0028 for $\tilde{R}<0.1$ and 0.0566 for $\tilde{R}>0.1$). The completeness is $80.7\%$, just slightly lower than with the real dust attenuation, and the precision is $78.3\%$, just lower than with the real $\rm{A_{1500}}$ and better than with \textsc{Cigale}'s $\rm{A_{1500}}$. The interpretation of those finding is the same as in Sect. \ref{sec:cigale}.

\begin{figure}
  \resizebox{\hsize}{!}{\includegraphics{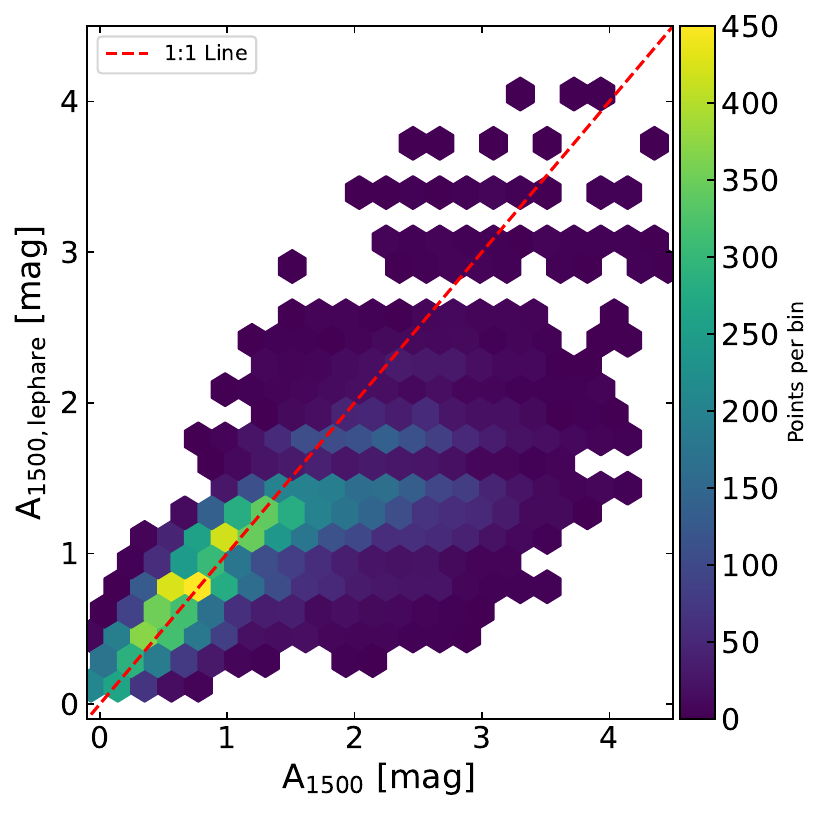}}
  \caption{Dust attenuation at $1500 \angstrom$ predicted by \textsc{lephare} compared to the true attenuation.}
  \label{fig:A1500_lephare}
\end{figure}

\end{appendix}

\end{document}